\documentclass[referee]{raa}
\usepackage{xcolor}
\usepackage{lineno}
\usepackage[utf8]{inputenc}
\usepackage{changes}
\usepackage{natbib}
\usepackage{mathptmx} 
\usepackage{booktabs}
\usepackage{siunitx}
\usepackage[flushleft]{threeparttable}
\usepackage{rotating}

\usepackage{graphicx,times}
\usepackage{amssymb,amsmath}
\bibpunct{(}{)}{;}{a}{}{,}

\usepackage[pagebackref=true]{hyperref}

\DeclareUnicodeCharacter{2212}{-}
\begin{document}

\title{The X-ray Variability of the Ultraluminous X-ray Sources in the NGC~4631  galaxy}

\author{Jiashun Chen\inst{1}, Jianfeng Wu\inst{1}, Zikun Lin\inst{1}}

\institute{Department of Astronomy, Xiamen University, Xiamen, Fujian 361005, China; {\it wujianfeng@xmu.edu.cn}}

\abstract{We perform a systematic study on the long-term X-ray variability for the five ultraluminous X-ray sources (ULXs) in the NGC 4631 galaxy (X1--X5), using \textit{Chandra}, \textit{XMM-Newton}, and \textit{Swift} observations covering a 24-year span. Light curves for the five ULXs are presented, while X-ray spectra were modeled for observations with sufficient counts. The normalized excess variance and structure function are utilized to study the X-ray variability behavior of the ULXs. The normalized excess variance is anti-correlated with average X-ray luminosity for three ULXs, indicating that objects with higher average luminosity  tend to exhibit relatively lower amplitude of variability. The structure function values  increases with time interval in two sources, showing that flux variations become more significant for longer  timescales. These trends are also found in the X-ray variability of active galactic nuclei (AGNs). The similarity between ULXs and AGNs, if confirmed for a larger sample of sources, possibly indicates similar underlying physical mechanisms for their X-ray variability. 
}

\authorrunning{J. Chen et al.}            
\titlerunning{X-ray Variability of NGC 4631 ULXs}  
\maketitle



\section{Introduction} \label{sec:intro}

Ultraluminous X-ray sources (ULXs) are off-nuclear, point-like high-energy objects whose X-ray luminosities exceed the isotropic Eddington luminosity of a typical stellar-mass ($\sim10\,M_{\odot}$) black hole \citep[e.g.,][]{Feng_2011, Kaaret_2017}. Since their discovery, ULXs have remained a key topic in high-energy astrophysics, yet their physical nature is still debated. Two main scenarios have been proposed to explain their extreme luminosities. In the first, the compact object---either a stellar-mass black hole or a neutron star---undergoes super-Eddington accretion, where geometrical beaming and radiation-driven winds from a thick accretion disk enhance the apparent luminosity beyond the classical Eddington limit \citep[e.g.,][]{Gladstone_2009,Walton_2014,Salvaggio_2022}. In the second, the accretor is an intermediate-mass black hole (IMBH; $10^{2}$–$10^{5}\,M_{\odot}$) radiating at a sub-Eddington rate \citep[e.g.,][]{Farrell_2009,Webb_2012}. Distinguishing between these two scenarios is crucial for understanding the black hole mass distribution, the final stages of massive stellar evolution, and the physics of extreme accretion.

During the accretion process, the emitted X-ray luminosity is expected to vary over time (e.g., \citealt{prokhorenko2024xrayvariabilitysdssquasars}). Studies of X-ray variability and spectral properties in other classes of sources, i.e. Galactic X-ray binaries and active galactic nuclei (AGNs), have provided  key constraints on the geometry and physical conditions of the accretion disk and corona, yielding estimates of fundamental black hole parameters such as mass, spin, and accretion rate. One of the most commonly used quantitative measures of variability is the normalized excess variance ($\sigma_{\mathrm{rms}}^{2}$), which describes the fractional amplitude of flux variations relative to the mean flux. Compared with power spectral density analysis, $\sigma_{\mathrm{rms}}^{2}$ has the advantage of being applicable to fainter or sparsely sampled sources, since it does not require continuous, high time–resolution monitoring \citep{Gonz_lez_Mart_n_2011}. In AGNs, $\sigma_{\mathrm{rms}}^{2}$ has been found to be anti-correlated with  X-ray luminosity   \citep{nandra1997dependence,turner1999x}. Applying this approach to ULXs, \citet{Gonz_lez_Mart_n_2011} reported that in the 2–10~keV band, the variability amplitude decreases with increasing luminosity, indicating a similar anti-correlation between variability and flux. Another used approach is the structure function (SF) method, which directly characterizes how variability evolves with timescale. Studies of AGNs have revealed that the variability amplitude increases with time separation and is inversely correlated with black hole mass, accretion rate, and X-ray luminosity\citep{prokhorenko2024xrayvariabilitysdssquasars}. These  approaches can be equally useful in probing the temporal behavior of ULXs.

The Whale Galaxy, NGC~4631, is viewed nearly edge-on and classified as a late-type starburst galaxy at a distance of 7.35~Mpc \citep{tully2013cosmicflows}. \citet{soria2009different} performed detailed analyses of five bright X-ray sources (X1--X5) using early \textit{Chandra} and \textit{XMM–Newton} data taken in 2000--2002, revealing a variety of flux and spectral properties—including supersoft, heavily absorbed, and persistent ULXs. However, the limited data then available precluded a systematic investigation of their long-term variability.
\citet{Guo__2023} combined  the astrometry from the \textit{Hubble Space Telescope} (HST) and the \textit{Chandra} X-ray Observatory data to precisely determine the positions of the five sources X1--X5 in NGC~4631 (see Table~\ref{tab:zb}). They investigated the stellar and gas environment of the sources, and found a new bubble nebula in the H$\alpha$ and [O~{\sc iii}] bands around the source X4 using the deep optical imaging by the Canada–France–Hawaii Telescope (CFHT). More recently, three additional ULXs (X6, X7, and X8)   have been identified in NGC 4631 with new \textit{XMM–Newton} data taken in 2025 \citep{Ducci_2025,Allak_2025}, one of which (X8) is proved to be a pulsar ULX with a spin period of 9.67~s \citep{Ducci_2025}.

\begin{table}[ht]
\small
\centering
\caption{Coordinates of the Five X-Ray Sources in NGC 4631 \label{tab:zb}}
\begin{tabular}{ccccc}
\hline
\textbf{Source ID} & &\textbf{R.A. } & & \textbf{Decl. }  \\
& & \textbf{(J2000)} & &  \textbf{(J2000)} \\
\hline
\ X1 & &12 42 15.99& & +32 32 49.47   \\
\ X2 && 12 42 11.12 && +32 32 35.63  \\
\ X3 && 12 42 06.13 && +32 32 46.43   \\
\ X4 & &12 41 57.42& & +32 32 02.79  \\
\ X5 && 12 41 55.57 && +32 32 16.77   \\
\hline
\end{tabular}
\end{table}



With the greatly increased number of X-ray observations now available, in this work we perform a systematic long-term variability study for the five ULXs  in NGC 4631, labeled as X1–X5, using the 37 archival X-ray observations  of NGC 4631 taken between 2000 and 2024. By constructing light curves, calculating hardness ratios, measuring normalized excess variances and the structure function, we aim to characterize the temporal behavior of these ULXs comprehensively and compare their variability patterns with those observed in AGNs.

\section{Data reduction} \label{sec:data}

The publicly available X-ray data of NGC 4631 from the \textit{Chandra} \citep{weisskopf2002overview}, \textit{XMM–Newton} \citep{jansen2001xmm}, and \textit{Swift} \citep{gehrels2004swift} missions were retrieved using the HEASARC \verb+Browse+ interface. \footnote{https://heasarc.gsfc.nasa.gov/cgi-bin/W3Browse/w3browse.pl} The datasets of 37 observations in total were obtained, including 12 from \textit{Chandra}, 2 from \textit{XMM-Newton}, \footnote{Another archival \textit{XMM-Newton} observation (ID 0890710201) in 2021 contains no usable data.} and 23 from \textit{Swift} (see Table~\ref{tab:obslog} for the observation log). 

\begin{table}[ht]
\small
\centering
\caption{X-ray Observation Log of NGC 4631 \label{tab:obslog}}
\begin{tabular}{lcccc}
\hline
\textbf{Mission} & \textbf{Observation ID} & \textbf{Date (UT)} & \textbf{Exposure Time (sec)} & \textbf{Detected sources}\\
 &  & & & \textbf{(0.5–8.0~keV)}\\
\hline
\textit{Chandra} & 797 & 2000-04-16 & 59970 & X2-X5\\
                 & 25777 & 2022-01-22 & 29420 &X2-X5 \\
                 & 25220 & 2022-08-02 & 23082& X1-X5\\
                 & 26484 & 2022-08-02 & 19070 &X1-X5\\
                 & 26485 & 2022-08-05 & 21080 &X2-X5\\
                 & 26486 & 2022-08-06 & 15080 &X2-X5\\
                 & 26487 & 2022-08-07 & 15080 &X2-X5\\
                 & 25782 & 2023-01-29 & 31080 &X2-X5\\
                 & 25780 & 2023-06-16 & 12080 &X2,X3,X5\\
                 & 25779 & 2023-07-04 & 20080 &X2-X5\\
                 & 25778 & 2023-07-04 & 20080 &X2-X5\\
                 & 25781 & 2023-07-18 & 13780 &X2-X5\\
\hline
\textit{XMM-Newton} & 0110900201 & 2002-06-28 & 54813& X1-X5\\
 & 0890710101 & 2021-12-28 & 33000& X2-X5 \\
\hline
\textit{Swift} & 00082263001 & 2013-11-08 & 7365.7 &X2,X3,X5\\
               & 00082263002 & 2013-11-10 & 2346.4 &X2,X3,X5\\
               & 00082263003 & 2013-11-16 & 2923.5& X2,X3,X5\\
               & 00082263004 & 2013-11-18 & 908.1 &X2,X5\\
               & 00082263005 & 2013-11-20 & 3518.5& X2,X3,X5\\
               & 00082263006 & 2013-11-21 & 1991.3& X2,X3,X5\\
               & 00084441001 & 2014-10-24 & 1258.7&X2,X5 \\
               & 00084441003 & 2018-03-23 & 386.1 &\\
               & 00084441004 & 2018-03-26 & 127.9 &\\
               & 00084441005 & 2018-05-20 & 313.4& X3,X5\\
               & 00084441006 & 2018-05-22 & 137.9& \\
               & 00084441007 & 2018-05-23 & 200.6 &\\
               & 00084441008 & 2018-11-21 & 563.0 &X2\\
               & 00084441009 & 2020-03-14 & 190.6 &X5\\
               & 00084441010 & 2020-08-11 & 747.3 &X5\\
               & 00084441011 & 2020-11-10 & 907.6 &X2,X5\\
               & 00084441012 & 2020-11-11 & 813.0 &X2,X5\\
               & 00084441013 & 2020-11-14 & 829.9 &X5\\
               & 00084441014 & 2020-11-16 & 508.2 &X2,X5\\
               & 00084441015 & 2020-11-18 & 526.6 &X5\\
               & 00084441016 & 2020-11-22 & 611.8 &X2,X5\\
               & 00084441017 & 2020-11-26 & 453.0& \\
               & 00084441018 & 2021-02-13 & 1837.8 &X3\\
\hline
\end{tabular}
\end{table}

For each mission, we started with the level~1 data. The \textit{Chandra} observations were reprocessed with the CIAO (v4.13) \citep{fruscione2006ciao} script \texttt{chandra\_repro}.\footnote{https://cxc.cfa.harvard.edu/ciao} Images and exposure maps in three energy bands (full: 0.5–8.0~keV, soft: 0.5–2.0~keV, and hard: 2.0–8.0~keV) were generated using the \texttt{fluximage} script. Background flare filtering was performed using the light curve extracted with \texttt{dmextract}. For the two \textit{XMM-Newton} observations, the XMMSAS (v18.0) \citep{gabriel2004xmm} \footnote{https://www.cosmos.esa.int/web/xmm-newton/sas} was used for data reduction.  The MOS and PN event lists were generated using the standard SAS tasks \texttt{emchain} and \texttt{epchain}, respectively, and the \texttt{evselect} command was used to extract event files in the same three energy bands.For event filtering, standard selection criteria were applied: for MOS data, we selected events with PATTERN $\leq$ 12 and FLAG == 0, while for PN data we used PATTERN $\leq$ 4 and FLAG == 0. The \textit{Swift} data were first  reprocessed with the \texttt{xrtpipeline} script. The command-line tool \texttt{xselect} in the HEASoft (v6.27)\footnote{https://heasarc.gsfc.nasa.gov/docs/software/lheasoft} package was used to extract events and produce images in the same three energy ranges listed above. As a result, comparable event files were obtained across all three instruments with consistent energy ranges.


. Circular regions centered on the known source coordinates were adopted as the source regions, with radii chosen according to the instrument point spread function (PSF) and the off-axis angle so as to enclose a fixed fraction of the encircled energy. For the \textit{Swift}/XRT and \textit{Chandra}/ACIS data, source radii of approximately $50^{\prime\prime}$ and $3$--$5^{\prime\prime}$ were adopted, corresponding to about 90\% encircled energy, respectively. For the \textit{XMM--Newton} observations, a source radius of about $15^{\prime\prime}$ was used, enclosing approximately 70\% and 73\% of the PSF encircled energy for the MOS and PN detectors, respectively. Annular regions surrounding the sources were used to estimate the background, ensuring that no other sources were present within the annuli.

\section{Data Analysis} \label{sec:data}
\subsection{X-ray Photometry} \label{sec:3.1}

Since the coordinates of the ULXs are already known, we adopt the source detection method used in \citet{pu2020fractionxrayweakquasars}.
The background-subtracted and aperture-corrected net counts were obtained by scaling the background counts to the source extraction area. The binomial probability for each energy band was calculated following Equation~(2) in \citet{pu2020fractionxrayweakquasars}:  
\begin{equation}
\mathrm{P_B} = \sum_{X=S}^{N} \frac{N!}{X!(N-X)!} \, p^{X} (1-p)^{N-X},
\label{eq:P_B}
\end{equation}
where $S$ is the total number of counts in the source region, $B$ is the total number of background counts, $N = S + B$, and $p = 1/(1 + \mathrm{BACKSCAL})$, with $\mathrm{BACKSCAL}$ denoting the area ratio of the background to source regions. A source was considered detected in a given energy band if $P_B < 0.001$. To check the validity of the source detection results using this methodology, we also performed blind source search procedures, with \texttt{wavdetect} in CIAO or \texttt{eboxdetect} in XMMSAS. The results obtained from both methods were highly consistent.

For the successfully detected sources, hardness ratios ($HR$) were calculated from the counts in the soft ($S$) and hard ($H$) bands ($HR=H-S/H+S$, as shown in Figure~\ref{fig:lc5}). Assuming a power-law spectral model with Galactic absorption for each observation, we estimated the effective photon index $\Gamma_{\rm eff}$ corresponding to the $HR$ value.We used PIMMS\footnote{https://cxc.harvard.edu/toolkit/pimms.jsp} v4.15 to construct the $HR$--$\Gamma_{\rm eff}$ relation, by fixing the 0.5--2 keV count rate as normalization and varying $\Gamma$ to compute the corresponding 2--8 keV count rate, assuming $N_{\rm H} = 1.29 \times 10^{20}\,\mathrm{cm^{-2}}$, which represents the Galactic absorption along the line of sight of NGC 4631. We therefore obtained a grid of $HR$ and $\Gamma_{\rm eff}$ values. The observed $HR$ values were then converted to $\Gamma_{\rm eff}$ via interpolation. This procedure was performed separately for each instrument to account for differences in instrumental response. The exposure map was used to determine the effective exposure time, which is used to obtain the precise count rate. The X-ray flux in the full band (0.5-8.0~keV) was calculated using the derived photon index $\Gamma_{\rm eff}$, which was then converted into X-ray luminosity with the adopted galaxy distance. The X-ray photometry results are presented in Tables~\ref{tab:flux2}–\ref{tab:flux5}) for X2--X5, respectively. X1 is a supersoft source  , with almost all the counts are in the soft band,  preventing us to derive $\Gamma_{\rm eff}$  from the hardness ratio. The X-ray luminosity of X1 is obtained from spectral fitting  (see the Section~\ref{sec:3.2}). For X2, the X-ray counts in \textit{Chandra} and \textit{XMM-Newton} observations are mostly in the hard band across all epochs, likely due to strong absorptions. A reliable estimate of the effective photon index $\Gamma_{\rm eff}$ cannot be obtained from hardness ratios. In contrast, the Swift observations provide a more balanced soft and hard band coverage, allowing a meaningful derivation of $\Gamma_{\rm eff}$, albeit with larger uncertainties. Therefore, only \textit{Swift} photometry of X2 with the derived $\Gamma_{\rm eff}$ is presented in Table~\ref{tab:flux2}.

\begin{table*}[ht]
\centering
\caption{X-ray Photometry of X2\label{tab:flux2}}
\resizebox{\textwidth}{!}{%
\begin{tabular}{cccccccccc}
\hline
Mission & ObsID & Soft-band & Hard-band & Full-band &  $HR$&$\Gamma_{\rm eff}$ & Effective & $f_X$ & $L_X$ \\
& & Net Counts &Net Counts &Net Counts & && Exposure (sec) & & \\
\hline
Swift & 00082263001 & $39.1^{+7.3}_{-6.2}$ & $27.7^{+6.3}_{-5.2}$ & $66.8^{+9.2}_{-8.2}$ & $-0.17 \pm 0.10$&$1.37^{+0.27}_{-0.21}$ & 7051.6 & $4.21^{+0.58}_{-0.51}$ & $2.72^{+0.37}_{-0.33}$ \\
Swift & 00082263005 & $10.7^{+4.4}_{-3.2}$ & $12.7^{+4.7}_{-3.5}$ & $23.5^{+6.9}_{-4.8}$ & $0.09 \pm 0.21$&$0.92^{+0.56}_{-0.34}$ & 868.8  & $3.86^{+0.97}_{-0.79}$ & $2.50^{+0.63}_{-0.51}$ \\
Swift & 00082263006 & $10.2^{+4.3}_{-3.1}$ & $7.1^{+3.8}_{-2.6}$  & $17.3^{+5.2}_{-4.1}$ & $-0.18 \pm 0.12$&$1.37^{+0.74}_{-0.40}$ & 3392.0 & $4.11^{+1.24}_{-0.98}$ & $2.66^{+0.80}_{-0.63}$ \\
\hline
\end{tabular}}

\vspace{2mm}
\noindent\footnotesize
\tablecomments{1.00\textwidth}{The table lists, for each observation, the mission name, observation ID, the soft-band, hard-band, and full-band net counts,$HR$, the derived effective photon index $\Gamma_{\rm eff}$, the effective exposure time, the full-band X-ray flux $f_X$ (in units of $10^{-13}$ erg cm$^{-2}$ s$^{-1}$), and the corresponding luminosity $L_X$ (in units of $10^{39}$ erg s$^{-1}$).}
\end{table*}

\begin{table*}[ht]
\centering
\caption{photometry of X3\label{tab:flux3}}

\resizebox{\textwidth}{!}{%
\begin{tabular}{cccccccccc}
\hline
Mission & ObsID & Soft-band & Hard-band & Full-band & $HR$ & $\Gamma_{\rm eff}$ & Effective & $f_X$ & $L_X$ \\
& & Net Counts &Net Counts &Net Counts & && Exposure (sec) & & \\
\hline
Chandra & 797   & $295.6^{+17.2}_{-17.2}$ & $145.7^{+12.1}_{-12.1}$ & $434.7^{+20.8}_{-20.8}$ & $-0.34 \pm 0.06$ & $1.08^{+0.10}_{-0.08}$ & 53804.4 & $0.76^{+0.04}_{-0.04}$ & $0.49^{+0.02}_{-0.02}$ \\
Chandra & 25220 & $32.3^{+6.7}_{-5.7}$ & $42.5^{+7.6}_{-6.5}$ & $72.3^{+9.5}_{-8.5}$ & $0.14 \pm 0.19$ & $1.24^{+0.34}_{-0.26}$ & 22458.6 & $0.55^{+0.07}_{-0.07}$ & $0.36^{+0.05}_{-0.04}$ \\
Chandra & 25778 & $43.2^{+7.6}_{-6.5}$ & $33.6^{+6.9}_{-5.8}$ & $77.9^{+9.9}_{-8.8}$ & $-0.13 \pm 0.10$ & $1.86^{+0.33}_{-0.26}$ & 19775.8 & $0.61^{+0.08}_{-0.07}$ & $0.39^{+0.05}_{-0.04}$ \\
Chandra & 25779 & $36.7^{+7.1}_{-6.0}$ & $30.4^{+6.6}_{-5.5}$ & $65.1^{+9.1}_{-8.0}$ & $-0.09 \pm 0.11$ & $1.79^{+0.36}_{-0.29}$ & 19735.5 & $0.51^{+0.07}_{-0.06}$ & $0.33^{+0.05}_{-0.04}$ \\
Chandra & 25780 & $12.7^{+4.7}_{-3.5}$ & $16.4^{+5.1}_{-4.0}$ & $29.4^{+6.5}_{-5.4}$ & $0.13 \pm 0.18$ & $1.26^{+0.65}_{-0.41}$ & 11543.6 & $0.44^{+0.10}_{-0.08}$ & $0.28^{+0.06}_{-0.05}$ \\
Chandra & 25781 & $24.1^{+6.0}_{-4.9}$ & $22.7^{+5.8}_{-4.7}$ & $46.7^{+7.9}_{-6.8}$ & $-0.03 \pm 0.12$ & $1.64^{+0.45}_{-0.33}$ & 13331.8 & $0.55^{+0.09}_{-0.08}$ & $0.36^{+0.06}_{-0.05}$ \\
Chandra & 25782 & $45.1^{+7.8}_{-6.7}$ & $44.6^{+7.7}_{-6.7}$ & $89.4^{+10.5}_{-9.4}$ & $0.00 \pm 0.11$ & $1.58^{+0.30}_{-0.24}$ & 30470.0 & $0.47^{+0.06}_{-0.05}$ & $0.30^{+0.04}_{-0.03}$ \\
Chandra & 26484 & $26.0^{+6.2}_{-5.1}$ & $28.7^{+6.4}_{-5.3}$ & $55.9^{+8.5}_{-7.4}$ & $0.05 \pm 0.13$ & $1.45^{+0.42}_{-0.30}$ & 18732.1 & $0.49^{+0.07}_{-0.07}$ & $0.32^{+0.05}_{-0.04}$ \\
Chandra & 26485 & $20.8^{+5.6}_{-4.5}$ & $36.6^{+7.1}_{-6.0}$ & $57.2^{+8.6}_{-7.5}$ & $0.27 \pm 0.12$ & $0.91^{+0.42}_{-0.30}$ & 20700.3 & $0.53^{+0.08}_{-0.07}$ & $0.34^{+0.05}_{-0.04}$ \\
Chandra & 26486 & $18.0^{+5.3}_{-4.2}$ & $25.8^{+6.1}_{-5.0}$ & $44.6^{+7.7}_{-6.7}$ & $0.18 \pm 0.14$ & $1.15^{+0.49}_{-0.34}$ & 14766.2 & $0.53^{+0.09}_{-0.08}$ & $0.34^{+0.06}_{-0.05}$ \\
Chandra & 26487 & $27.6^{+6.3}_{-5.2}$ & $22.7^{+5.8}_{-4.7}$ & $49.3^{+8.1}_{-7.0}$ & $-0.10 \pm 0.12$ & $1.80^{+0.43}_{-0.33}$ & 14732.4 & $0.52^{+0.09}_{-0.07}$ & $0.33^{+0.06}_{-0.05}$ \\
\hline
XMM(MOS1) & 110900201 & $198.7^{+14.1}_{-14.1}$ & $71.2^{+9.5}_{-8.4}$ & $270.2^{+16.4}_{-16.4}$ & $0.47 \pm 0.04$ & $1.85^{+0.14}_{-0.12}$ & 46671.4 & $0.51^{+0.03}_{-0.03}$ & $0.33^{+0.02}_{-0.02}$ \\
XMM(MOS2) & 110900201 & $172.4^{+13.1}_{-13.1}$ & $103.1^{+10.2}_{-10.2}$ & $275.6^{+16.6}_{-16.6}$ & $-0.25 \pm 0.05$ & $1.39^{+0.12}_{-0.10}$ & 47779.1 & $0.63^{+0.04}_{-0.04}$ & $0.41^{+0.03}_{-0.03}$ \\
XMM(PN) & 110900201 & $438.1^{+20.9}_{-20.9}$ & $181.3^{+13.5}_{-13.5}$ & $619.4^{+24.9}_{-24.9}$ & $-0.38 \pm 0.03$ & $1.48^{+0.07}_{-0.07}$ & 41236.6 & $0.48^{+0.02}_{-0.02}$ & $0.31^{+0.01}_{-0.01}$ \\
XMM(MOS1) & 890710101 & $107.3^{+10.4}_{-10.4}$ & $98.9^{+10.9}_{-9.9}$ & $206.1^{+14.4}_{-14.4}$ & $-0.04 \pm 0.07$ & $1.01^{+0.14}_{-0.12}$ & 23273.0 & $1.20^{+0.08}_{-0.08}$ & $0.78^{+0.05}_{-0.05}$ \\
XMM(MOS2) & 890710101 & $122.5^{+11.1}_{-11.1}$ & $69.5^{+9.4}_{-8.3}$ & $192.0^{+13.9}_{-13.9}$ & $0.28 \pm 0.07$ & $1.44^{+0.15}_{-0.13}$ & 25258.6 & $0.82^{+0.06}_{-0.06}$ & $0.53^{+0.04}_{-0.04}$ \\
XMM(PN) & 890710101 & $208.8^{+14.4}_{-14.4}$ & $93.4^{+10.6}_{-9.7}$ & $302.2^{+17.4}_{-17.4}$ & $-0.38 \pm 0.05$ & $1.44^{+0.11}_{-0.09}$ & 12320.8 & $0.79^{+0.05}_{-0.05}$ & $0.51^{+0.03}_{-0.03}$ \\
\hline
Swift & 00082263005 & $13.9^{+4.8}_{-3.7}$ & $12.2^{+4.5}_{-3.4}$ & $26.0^{+6.2}_{-5.1}$ & $0.07 \pm 0.21$ & $1.18^{+0.52}_{-0.32}$ & 3291.5 & $3.86^{+0.91}_{-0.75}$ & $2.50^{+0.59}_{-0.49}$ \\
\hline
\end{tabular}}

\vspace{2mm}
\noindent\footnotesize
\tablecomments{1.00\textwidth}{The columns have the same definitions as those in Table~\ref{tab:flux2}.}
\end{table*}

\begin{table*}[ht]
\centering
\caption{Photometry of X4\label{tab:flux4}}
\resizebox{\textwidth}{!}{%
\begin{tabular}{cccccccccc}
\hline
Mission & ObsID & Soft-band & Hard-band & Full-band & $HR$ & $\Gamma_{\rm eff}$ & Effective & $f_X$ & $L_X$ \\
& & Net Counts &Net Counts &Net Counts & && Exposure (sec) & & \\
\hline
Chandra & 797   & $81.1^{+10.0}_{-9.0}$ & $14.1^{+4.8}_{-3.7}$ & $92.2^{+10.6}_{-9.6}$ & $-0.70 \pm 0.04$ & $1.94^{+0.33}_{-0.23}$ & 56893.9 & $0.08^{+0.01}_{-0.01}$ & $0.05^{+0.01}_{-0.01}$ \\
Chandra & 25777 & $94.8^{+10.7}_{-9.7}$ & $63.0^{+9.0}_{-7.9}$ & $157.1^{+12.5}_{-12.5}$ & $-0.20 \pm 0.07$ & $2.04^{+0.22}_{-0.18}$ & 28144.9 & $0.85^{+0.07}_{-0.07}$ & $0.55^{+0.04}_{-0.04}$ \\
Chandra & 25778 & $145.2^{+12.0}_{-12.0}$ & $106.0^{+10.3}_{-10.3}$ & $249.3^{+15.8}_{-15.8}$ & $-0.16 \pm 0.05$ & $1.93^{+0.16}_{-0.14}$ & 19768.3 & $1.93^{+0.12}_{-0.12}$ & $1.25^{+0.08}_{-0.08}$ \\
Chandra & 25779 & $20.2^{+5.6}_{-4.5}$ & $26.1^{+6.2}_{-5.1}$ & $45.2^{+7.8}_{-6.7}$ & $0.13 \pm 0.13$ & $1.26^{+0.47}_{-0.33}$ & 19726.7 & $0.39^{+0.07}_{-0.06}$ & $0.25^{+0.04}_{-0.04}$ \\
Chandra & 25781 & $74.1^{+9.6}_{-8.6}$ & $69.1^{+9.4}_{-8.3}$ & $140.7^{+11.9}_{-11.9}$ & $-0.03 \pm 0.07$ & $1.65^{+0.23}_{-0.19}$ & 13347.6 & $1.66^{+0.14}_{-0.14}$ & $1.07^{+0.09}_{-0.09}$ \\
Chandra & 25782 & $36.6^{+7.1}_{-6.0}$ & $31.1^{+6.6}_{-5.5}$ & $67.6^{+9.3}_{-8.2}$ & $-0.08 \pm 0.09$ & $1.76^{+0.35}_{-0.28}$ & 30635.9 & $0.34^{+0.05}_{-0.04}$ & $0.22^{+0.03}_{-0.03}$ \\
Chandra & 26484 & $24.7^{+6.0}_{-4.9}$ & $39.2^{+7.3}_{-6.2}$ & $63.7^{+9.0}_{-8.0}$ & $0.23 \pm 0.10$ & $1.04^{+0.38}_{-0.28}$ & 18411.3 & $0.63^{+0.09}_{-0.08}$ & $0.41^{+0.06}_{-0.05}$ \\
Chandra & 26485 & $9.6^{+4.2}_{-3.0}$ & $8.2^{+4.0}_{-2.8}$ & $17.1^{+5.2}_{-4.1}$ & $-0.08 \pm 0.17$ & $1.74^{+0.96}_{-0.53}$ & 20757.6 & $0.13^{+0.04}_{-0.03}$ & $0.08^{+0.03}_{-0.02}$ \\
Chandra & 26486 & $14.9^{+4.9}_{-3.8}$ & $15.2^{+5.0}_{-3.9}$ & $29.7^{+6.5}_{-5.4}$ & $0.01 \pm 0.15$ & $1.53^{+0.63}_{-0.40}$ & 14835.7 & $0.32^{+0.07}_{-0.06}$ & $0.21^{+0.05}_{-0.04}$ \\
Chandra & 26487 & $23.5^{+5.9}_{-4.8}$ & $31.1^{+6.6}_{-5.6}$ & $55.3^{+8.5}_{-7.4}$ & $0.14 \pm 0.11$ & $1.23^{+0.42}_{-0.30}$ & 13257.7 & $0.72^{+0.11}_{-0.10}$ & $0.47^{+0.07}_{-0.06}$ \\
\hline
XMM(MOS1) & 110900201 & $500.5^{+22.4}_{-22.4}$ & $259.0^{+16.1}_{-16.1}$ & $759.5^{+27.6}_{-27.6}$ & $-0.32 \pm 0.03$ & $1.52^{+0.07}_{-0.06}$ & 46932.1 & $1.67^{+0.06}_{-0.06}$ & $1.08^{+0.04}_{-0.04}$ \\
XMM(MOS2) & 110900201 & $516.1^{+22.7}_{-22.7}$ & $247.8^{+15.7}_{-15.7}$ & $763.4^{+27.6}_{-27.6}$ & $-0.35 \pm 0.03$ & $1.59^{+0.07}_{-0.07}$ & 47406.1 & $1.60^{+0.06}_{-0.06}$ & $1.04^{+0.04}_{-0.04}$ \\
XMM(PN) & 110900201 & $1162.9^{+34.1}_{-34.1}$ & $496.7^{+22.3}_{-22.3}$ & $1659.6^{+40.7}_{-40.7}$ & $-0.40 \pm 0.02$ & $1.45^{+0.04}_{-0.04}$ & 41333.8 & $1.30^{+0.03}_{-0.03}$ & $0.84^{+0.02}_{-0.02}$ \\
XMM(MOS1) & 890710101 & $153.1^{+12.4}_{-12.4}$ & $97.3^{+10.9}_{-9.8}$ & $250.4^{+15.8}_{-15.8}$ & $-0.22 \pm 0.05$ & $1.34^{+0.13}_{-0.11}$ & 24349.3 & $1.16^{+0.07}_{-0.07}$ & $0.75^{+0.05}_{-0.04}$ \\
XMM(MOS2) & 890710101 & $164.6^{+12.8}_{-12.8}$ & $109.0^{+10.4}_{-10.4}$ & $273.6^{+16.5}_{-16.5}$ & $-0.20 \pm 0.05$ & $1.31^{+0.12}_{-0.10}$ & 27112.2 & $1.16^{+0.07}_{-0.07}$ & $0.75^{+0.05}_{-0.05}$ \\
XMM(PN) & 890710101 & $267.1^{+16.3}_{-16.3}$ & $110.5^{+10.5}_{-10.5}$ & $377.6^{+19.4}_{-19.4}$ & $-0.42 \pm 0.04$ & $1.50^{+0.09}_{-0.08}$ & 7994.2  & $1.47^{+0.08}_{-0.08}$ & $0.95^{+0.05}_{-0.05}$ \\
\hline
\end{tabular}}

\vspace{2mm}
\noindent\footnotesize
\tablecomments{1.00\textwidth}{The columns have the same definitions as those in Table~\ref{tab:flux2}.}
\end{table*}

\begin{table*}[ht]
\centering
\caption{Photometry of X5\label{tab:flux5}}
\resizebox{\textwidth}{!}{%
\begin{tabular}{cccccccccc}
\hline
Mission & ObsID & Soft-band & Hard-band & Full-band & $HR$ & $\Gamma_{\rm eff}$ & Effective & $f_X$ & $L_X$ \\
& & Net Counts &Net Counts &Net Counts & && Exposure (sec) & & \\
\hline
Chandra & 797   & $2573.7^{+50.7}_{-50.7}$ & $1001.5^{+31.6}_{-31.6}$ & $3552.2^{+59.6}_{-59.6}$ & $-0.44^{+0.01}_{-0.01}$ & $1.28^{+0.03}_{-0.03}$ & 57207.6 & $5.01^{+0.08}_{-0.08}$ & $3.24^{+0.05}_{-0.05}$ \\
Chandra & 25220 & $153.5^{+12.4}_{-12.4}$ & $145.3^{+12.1}_{-12.1}$ & $296.0^{+17.2}_{-17.2}$ & $-0.03^{+0.07}_{-0.07}$ & $1.63^{+0.15}_{-0.14}$ & 22014.4 & $2.12^{+0.12}_{-0.12}$ & $1.37^{+0.08}_{-0.08}$ \\
Chandra & 25777 & $808.1^{+28.4}_{-28.4}$ & $904.1^{+30.1}_{-30.1}$ & $1701.4^{+41.2}_{-41.2}$ & $0.06^{+0.03}_{-0.03}$ & $1.43^{+0.05}_{-0.05}$ & 28179.8 & $9.91^{+0.24}_{-0.24}$ & $6.41^{+0.16}_{-0.16}$ \\
Chandra & 25778 & $185.7^{+13.6}_{-13.6}$ & $176.2^{+13.3}_{-13.3}$ & $359.4^{+19.0}_{-19.0}$ & $-0.03^{+0.05}_{-0.05}$ & $1.63^{+0.13}_{-0.12}$ & 19708.8 & $2.88^{+0.15}_{-0.15}$ & $1.86^{+0.10}_{-0.10}$ \\
Chandra & 25779 & $201.1^{+14.2}_{-14.2}$ & $217.1^{+14.7}_{-14.7}$ & $416.9^{+20.4}_{-20.4}$ & $0.04^{+0.05}_{-0.05}$ & $1.47^{+0.12}_{-0.11}$ & 19747.7 & $3.44^{+0.17}_{-0.17}$ & $2.22^{+0.11}_{-0.11}$ \\
Chandra & 25780 & $99.3^{+11.0}_{-9.9}$ & $90.9^{+10.5}_{-9.5}$ & $187.7^{+13.7}_{-13.7}$ & $-0.04^{+0.07}_{-0.07}$ & $1.67^{+0.19}_{-0.17}$ & 11900.3 & $2.48^{+0.18}_{-0.18}$ & $1.60^{+0.12}_{-0.12}$ \\
Chandra & 25781 & $130.8^{+11.4}_{-11.4}$ & $108.6^{+10.4}_{-10.4}$ & $237.2^{+15.4}_{-15.4}$ & $-0.09^{+0.06}_{-0.06}$ & $1.79^{+0.16}_{-0.15}$ & 13377.9 & $2.75^{+0.18}_{-0.18}$ & $1.77^{+0.11}_{-0.11}$ \\
Chandra & 25782 & $796.1^{+28.2}_{-28.2}$ & $1137.4^{+33.7}_{-33.7}$ & $1924.8^{+43.9}_{-43.9}$ & $0.18^{+0.02}_{-0.02}$ & $1.15^{+0.05}_{-0.05}$ & 30578.4 & $11.1^{+0.3}_{-0.3}$ & $7.19^{+0.16}_{-0.17}$ \\
Chandra & 26484 & $107.3^{+10.4}_{-10.4}$ & $119.2^{+10.9}_{-10.9}$ & $226.6^{+15.1}_{-15.1}$ & $0.05^{+0.07}_{-0.07}$ & $1.44^{+0.17}_{-0.15}$ & 18647.3 & $1.99^{+0.13}_{-0.13}$ & $1.29^{+0.09}_{-0.09}$ \\
Chandra & 26485 & $142.6^{+11.9}_{-11.9}$ & $174.9^{+13.2}_{-13.2}$ & $316.5^{+17.8}_{-17.8}$ & $0.10^{+0.06}_{-0.06}$ & $1.32^{+0.14}_{-0.13}$ & 20565.8 & $2.59^{+0.15}_{-0.14}$ & $1.67^{+0.09}_{-0.09}$ \\
Chandra & 26486 & $122.6^{+11.1}_{-11.1}$ & $112.6^{+10.6}_{-10.6}$ & $232.0^{+15.2}_{-15.2}$ & $-0.04^{+0.07}_{-0.07}$ & $1.67^{+0.17}_{-0.15}$ & 14635.6 & $2.49^{+0.16}_{-0.16}$ & $1.61^{+0.11}_{-0.11}$ \\
Chandra & 26487 & $105.8^{+10.3}_{-10.3}$ & $104.8^{+10.2}_{-10.2}$ & $211.7^{+14.5}_{-14.5}$ & $-0.00^{+0.07}_{-0.07}$ & $1.57^{+0.18}_{-0.15}$ & 12846.8 & $2.63^{+0.18}_{-0.18}$ & $1.70^{+0.12}_{-0.12}$ \\
\hline
XMM(MOS1) & 110900201 & $1480.8^{+38.5}_{-38.5}$ & $754.3^{+27.5}_{-27.5}$ & $2233.7^{+47.3}_{-47.3}$ & $-0.33^{+0.02}_{-0.02}$ & $1.54^{+0.04}_{-0.04}$ & 46763.7 & $4.88^{+0.10}_{-0.10}$ & $3.15^{+0.07}_{-0.07}$ \\
XMM(MOS2) & 110900201 & $1485.3^{+38.5}_{-38.5}$ & $649.9^{+25.5}_{-25.5}$ & $2133.8^{+46.2}_{-46.2}$ & $-0.39^{+0.02}_{-0.02}$ & $1.67^{+0.04}_{-0.04}$ & 46931.1 & $4.35^{+0.09}_{-0.09}$ & $2.81^{+0.06}_{-0.06}$ \\
XMM(PN) & 110900201 & $3653.2^{+60.4}_{-60.4}$ & $1441.5^{+38.0}_{-38.0}$ & $5093.3^{+71.4}_{-71.4}$ & $-0.43^{+0.01}_{-0.01}$ & $1.52^{+0.02}_{-0.02}$ & 41163.1 & $3.86^{+0.05}_{-0.05}$ & $2.49^{+0.03}_{-0.03}$ \\
XMM(MOS1) & 890710101 & $480.3^{+21.9}_{-21.9}$ & $200.8^{+14.2}_{-14.2}$ & $681.3^{+26.1}_{-26.1}$ & $-0.41^{+0.04}_{-0.04}$ & $1.71^{+0.08}_{-0.07}$ & 23100.1 & $2.77^{+0.11}_{-0.11}$ & $1.79^{+0.07}_{-0.07}$ \\
XMM(MOS2) & 890710101 & $624.7^{+25.0}_{-25.0}$ & $251.2^{+15.9}_{-15.9}$ & $875.9^{+29.6}_{-29.6}$ & $-0.43^{+0.03}_{-0.03}$ & $1.74^{+0.07}_{-0.06}$ & 27145.6 & $2.98^{+0.10}_{-0.10}$ & $1.92^{+0.07}_{-0.07}$ \\
XMM(PN) & 890710101 & $770.9^{+27.8}_{-27.8}$ & $272.1^{+16.5}_{-16.5}$ & $1043.0^{+32.3}_{-32.3}$ & $-0.48^{+0.03}_{-0.03}$ & $1.62^{+0.06}_{-0.05}$ & 12164.0 & $2.46^{+0.08}_{-0.08}$ & $1.59^{+0.05}_{-0.05}$ \\
\hline
Swift & 82263001 & $75.8^{+9.7}_{-8.7}$ & $52.9^{+8.3}_{-7.3}$ & $128.7^{+10.3}_{-11.3}$ & $-0.18^{+0.10}_{-0.10}$ & $1.38^{+0.18}_{-0.15}$ & 6218.5 & $9.14^{+0.81}_{-0.81}$ & $5.91^{+0.52}_{-0.52}$ \\
Swift & 82263002 & $30.3^{+6.6}_{-5.5}$ & $18.4^{+5.4}_{-4.3}$ & $48.6^{+8.0}_{-7.0}$ & $-0.24^{+0.18}_{-0.18}$ & $1.50^{+0.35}_{-0.24}$ & 1919.6 & $10.6^{+1.7}_{-1.5}$ & $6.82^{+1.12}_{-0.98}$ \\
Swift & 82263003 & $27.4^{+6.3}_{-5.2}$ & $22.8^{+5.8}_{-4.7}$ & $50.1^{+8.1}_{-7.1}$ & $-0.09^{+0.17}_{-0.17}$ & $1.22^{+0.33}_{-0.24}$ & 2607.5 & $9.17^{+1.49}_{-1.29}$ & $5.93^{+0.96}_{-0.83}$ \\
Swift & 82263004 & $7.3^{+3.8}_{-2.6}$ & $16.3^{+5.1}_{-3.0}$ & $23.7^{+5.9}_{-4.8}$ & $0.38^{+0.32}_{-0.32}$ & $0.37^{+0.64}_{-0.36}$ & 795.5 & $21.7^{+5.4}_{-4.4}$ & $14.0^{+3.5}_{-2.9}$ \\
Swift & 82263005 & $40.4^{+7.4}_{-6.3}$ & $16.7^{+5.2}_{-4.0}$ & $57.1^{+8.6}_{-7.5}$ & $-0.41^{+0.14}_{-0.14}$ & $1.83^{+0.34}_{-0.24}$ & 3034.5 & $6.72^{+1.01}_{-0.89}$ & $4.35^{+0.65}_{-0.57}$ \\
Swift & 82263006 & $17.1^{+5.2}_{-4.0}$ & $11.5^{+4.5}_{-3.3}$ & $28.6^{+6.4}_{-5.3}$ & $-0.20^{+0.22}_{-0.22}$ & $1.41^{+0.50}_{-0.31}$ & 1756.8 & $7.05^{+1.58}_{-1.31}$ & $4.56^{+1.02}_{-0.85}$ \\
Swift & 84441001 & $13.3^{+4.7}_{-3.6}$ & $8.1^{+4.0}_{-2.8}$ & $21.4^{+5.7}_{-4.6}$ & $-0.24^{+0.25}_{-0.25}$ & $1.49^{+0.64}_{-0.36}$ & 1146.5 & $7.79^{+2.08}_{-1.67}$ & $5.04^{+1.34}_{-1.08}$ \\
Swift & 84441008 & $12.0^{+4.6}_{-3.4}$ & $5.5^{+3.5}_{-2.3}$ & $17.4^{+5.3}_{-4.1}$ & $-0.37^{+0.32}_{-0.32}$ & $1.74^{+0.82}_{-0.42}$ & 559.4 & $11.6^{+3.5}_{-2.7}$ & $7.48^{+2.26}_{-1.77}$ \\
Swift & 84441011 & $7.9^{+3.9}_{-2.7}$ & $11.0^{+4.4}_{-3.3}$ & $18.9^{+5.4}_{-4.3}$ & $0.16^{+0.28}_{-0.28}$ & $0.77^{+0.69}_{-0.38}$ & 855.0 & $13.3^{+3.8}_{-3.0}$ & $8.57^{+2.46}_{-1.95}$ \\
Swift & 84441012 & $13.2^{+4.7}_{-3.6}$ & $11.9^{+4.5}_{-3.4}$ & $25.0^{+6.1}_{-5.0}$ & $-0.05^{+0.21}_{-0.21}$ & $1.16^{+0.53}_{-0.33}$ & 682.0 & $18.1^{+4.4}_{-3.6}$ & $11.7^{+2.8}_{-2.3}$ \\
Swift & 84441013 & $6.4^{+3.7}_{-2.5}$ & $5.3^{+3.4}_{-2.2}$ & $11.7^{+4.5}_{-3.4}$ & $-0.09^{+0.35}_{-0.35}$ & $1.23^{+1.09}_{-0.47}$ & 511.8 & $10.9^{+4.2}_{-3.1}$ & $7.03^{+2.72}_{-2.02}$ \\
Swift & 84441016 & $12.2^{+4.7}_{-3.5}$ & $5.3^{+3.4}_{-2.2}$ & $18.1^{+5.3}_{-4.2}$ & $-0.39^{+0.30}_{-0.30}$ & $1.83^{+0.81}_{-0.42}$ & 597.4 & $10.9^{+3.2}_{-2.5}$ & $7.01^{+2.06}_{-1.63}$ \\
\hline
\end{tabular}}

\vspace{2mm}
\noindent\footnotesize
\tablecomments{1.00\textwidth}{The columns have the same definitions as those in Table~\ref{tab:flux2}.}
\end{table*}

\subsection{X-ray Spectroscopy} \label{sec:3.2}

X-ray spectral modeling was performed using \textsc{XSPEC}(v12) implemented in the HEASoft package\citep{arnaud1996xspec}. \textit{Chandra} spectra and corresponding response files were extracted with CIAO/\texttt{specextract}. For \textit{XMM-Newton} data, the source and background spectra were extracted with the \texttt{evselect} command, while the response matrix and effective area files were generated using \texttt{rmfgen} and \texttt{arfgen}, respectively. Simultaneous spectra from the two MOS instruments were first combined using \texttt{epicspeccombine} before performing the fit. The PN spectra were modeled separately. For \textit{Swift}/XRT data,  the source and background spectral extractions, were carried out with \texttt{xselect}. The ancillary response files were generated using \texttt{xrtmkarf} and the standard Swift/XRT RMF from CALDB (v4.9.4) was used. For all three missions, the full band spectra in 0.5--8.0~keV were input to \textsc{XSPEC} for modeling. 

Depending on the photon statistics, two fitting strategies were adopted: for spectra with more than 100 detected counts, we used $\chi^2$ fitting with at least 10 counts per spectral bin. Increasing the counts per bin would be statistically preferable, but would leave insufficient bins for reliable spectral fitting. For those with fewer than 100 counts, Cash statistics ($Cstat$; \citealt{Cash_1979})  was applied. However, most of the spectral fittings with the Cash statistics could not provide meaningful constraints, except for a few cases that  will be mentioned in discussions below (Section~\ref{subsec:five_ulxs}). Therefore, we mainly present the spectral modeling results with the $\chi^2$ statistics.  We adopt the spectral model selection of \citet{soria2009different}, which we have re-tested on our expanded dataset and found to remain the most appropriate, as alternative models result in poorer fits.The absorption models wabs and phabs in XSPEC were used with their default abundance tables (\citet{morrison1983interstellar} for wabs and \citet{anders1989abundances} for phabs). The Galactic column density was fixed at $N_{\rm H} = 1.2 \times 10^{20}\,\mathrm{cm^{-2}}$, following and ensuring consistency with \citet{soria2009different}. The \textit{XMM-Newton}/MOS spectrum of X1 was fitted using the spectral model containing absorbed (both Galactic and intrinsic) blackbody and optically-thin plasma component with an absorption edge (\texttt{phabs*phabs*(bbody+raymond)*zedge}). The spectra of X2 and X3 were modeled with absorbed disk blackbody models (\texttt{wabs*wabs*diskbb}). The \textit{XMM-Newton}/MOS spectra of X4 were modeled with both optically-thin plasma and the disk blackbody plus power-law components (\texttt{wabs*(wabs*raymond + wabs*(powerlaw + diskbb))}). The spectra of X5 were fitted with absorbed power-law models (\texttt{wabs*wabs*powerlaw}). The PN spectra were not used in cases where the source fell on CCD gaps, or suffered from poor data quality. The goodness of fit was evaluated using the ratio of $\chi^2$  to degrees of freedom, with ideal values close to unity. From the best-fit models, we extracted the absorbed fluxes in the 0.5--8~keV band, which were then converted to X-ray luminosities. The corresponding best-fit parameters for all the spectral modeling with $\chi^2$ statistics are summarized in Tables~\ref{tab:1}–\ref{tab:5} for X1--X5, repsectively.


\begin{table}[ht]
\centering
\caption{XMM-Newton Spectral fitting data of X1\label{tab:1}}
\begin{tabular}{p{3.6cm} >{\centering\arraybackslash}p{3.6cm}}
\hline
\multicolumn{1}{l}{Parameter} & ObsID (110900201) \\
\hline
$N_{\mathrm{H, Gal}}$ & $(1.3 \times 10^{20})$ \\
$N_{\mathrm{H}}$ & $1.8^{+1.9}_{-1.5} \times 10^{21}$ \\
$kT_{\mathrm{rs}}$ (keV) & $0.46^{+0.20}_{-0.17}$ \\
$Z(Z_{\odot})$ & $(1.0)$ \\
$K_{\mathrm{rs}}$ & $8.1^{+2.2}_{-4.7} \times 10^{-6}$ \\
$kT_{\mathrm{bb}}$ (keV) & $0.09^{+0.03}_{-0.02}$ \\
$K_{\mathrm{bb}}$ & $4.0^{+4.0}_{-2.5} \times 10^{-6}$ \\
$E_{\mathrm{edge}}$ (keV) & $0.98^{+0.04}_{-0.06}$ \\
$\tau_{\mathrm{edge}}$ & $1.7^{+2.1}_{-0.9}$ \\
\hline
$\chi^2/\mathrm{dof}$ & $0.76(41.5/55)$ \\
$f_X$ & $3.88^{+0.13}_{-3.21} \times 10^{-14}$ \\
$L_X$ & $2.51^{+0.08}_{-2.08} \times 10^{38}$ \\
\hline
\end{tabular}

\vspace{2mm}
\noindent\footnotesize
The best-fit parameter values for the \textit{XMM-Newton} spectra of X1. 
Errors are 90\% confidence levels for one interesting parameter.
$N_{\mathrm{H,Gal}}$ and $N_{\mathrm{H}}$ are the Galactic and intrinsic absorption column density in units of $\mathrm{cm^{-2}}$, respectively. 
$kT_{\mathrm{rs}}$ and $kT_{\mathrm{bb}}$ are the temperatures of the \texttt{raymond-smith} and \texttt{blackbody} components in keV, while $K_{\mathrm{rs}}$ and $K_{\mathrm{bb}}$ are the normalizations of those two components, respectively. 
$E_{\mathrm{edge}}$ and $\tau_{\mathrm{edge}}$ are the energy and optical depth of the absorption edge,
while $Z(Z_{\odot})$ represents metallicity (fixed at 1.0),
$f_X$ is the observed full-band X-ray flux in units of $\mathrm{erg\,cm^{-2}\,s^{-1}}$,
while $L_X$ is the corresponding full-band X-ray luminosity in units of $\mathrm{erg\,s^{-1}}$. 
\end{table}

\begin{table*}[ht]
\centering
\caption{Spectral fitting data of X2\label{tab:2}}
\resizebox{\textwidth}{!}{%
\begin{tabular}{ccccccccc}
\hline
Mission & ObsID & $N_{\mathrm{H,Gal}}$ & $N_{\mathrm{H}}$ & $K_{\mathrm{dbb}}$ & $kT_{\mathrm{dbb}}$ & $f_X$ & $L_X$ & $\chi^2/\mathrm{dof}$ \\
\hline
Chandra  & 797       & 1.3 & $293.5^{+44.7}_{-39.1}$ & $7.21^{+6.43}_{-3.48}\times 10^{-3}$ & $1.40^{+0.21}_{-0.17}$ & $2.94^{+0.23}_{-0.87}$ & $1.90^{+0.15}_{-0.56}$ & 0.75(28.5/38) \\
\hline
XMM(MOS) & 110900201 & 1.3 & $272.3^{+27.1}_{-40.7}$ & $9.20^{+6.95}_{-3.92}\times 10^{-3}$ & $1.25^{+0.15}_{-0.13}$ & $2.87^{+0.12}_{-0.84}$ & $1.86^{+0.08}_{-0.54}$ & 1.17(57.4/49) \\
XMM(MOS) & 890710101 & 1.3 & $225.8^{+75.1}_{-57.3}$ & $12.20^{+19.27}_{-7.19}\times 10^{-3}$ & $1.11^{+0.22}_{-0.18}$ & $2.29^{+0.15}_{-1.68}$ & $1.48^{+0.10}_{-1.09}$ & 1.41(43.8/31) \\
XMM(PN)  & 890710101 & 1.3 & $96.3^{+57.3}_{-36.8}$  & $1.09^{+1.75}_{-0.65}\times 10^{-3}$ & $1.84^{+0.47}_{-0.36}$ & $2.32^{+0.15}_{-1.36}$ & $1.50^{+0.10}_{-0.88}$ & 1.18(55.3/47) \\
\hline
\end{tabular}}

\vspace{2mm}
\noindent\footnotesize
The X-ray spectral fitting results for X2. 
Errors are 90\% confidence levels for one interesting parameter.
$N_{\mathrm{H,Gal}}$ and $N_{\mathrm{H}}$ are the Galactic and intrinsic absorption column density in units of $10^{20}\,\mathrm{cm^{-2}}$, respectively. 
$K_{\mathrm{dbb}}$ and $kT_{\mathrm{dbb}}$ are the normalization and the inner disk temperature (in keV) of the \texttt{diskbb} model,
$f_X$ is the observed full-band X-ray flux in units of $10^{-13}\,\mathrm{erg\,cm^{-2}\,s^{-1}}$,
while $L_X$ is the corresponding full-band X-ray luminosity in units of $10^{39}\,\mathrm{erg\,s^{-1}}$. 
\end{table*}

\begin{table*}[ht]
\centering
\caption{Spectral fitting data of X3\label{tab:3}}
\resizebox{\textwidth}{!}{%
\begin{tabular}{ccccccccc}
\hline
Mission & obsID & $N_{\mathrm{H,Gal}}$ & $N_{\mathrm{H}}$ & $K_{\mathrm{dbb}}$ &  $kT_{\mathrm{dbb}}$ & $f_X$ & $L_X$ & $\chi^2/\mathrm{dof}$ \\
\hline
Chandra  & 797       & 1.3 & $25.9^{+9.6}_{-7.8}$ & $0.98^{+0.90}_{-0.48}\times 10^{-3}$ & $1.32^{+0.25}_{-0.20}$ & $5.23^{+0.40}_{-2.32}$ & $3.38^{+0.26}_{-1.50}$ & 0.84(28.6/34) \\
\hline
XMM(MOS) & 110900201 & 1.3 & $2.7^{+5.3}_{-2.7}$ & $1.23^{+1.99}_{-0.74}\times 10^{-3}$ & $1.18^{+0.33}_{-0.26}$ & $6.43^{+0.54}_{-4.22}$ & $4.15^{+0.40}_{-2.72}$ & 0.90(25.2/28) \\
XMM(PN)  & 110900201 & 1.3 & $17.9^{+9.6}_{-7.4}$ & $2.18^{+3.51}_{-1.26}\times 10^{-3}$ & $1.02^{+0.26}_{-0.23}$ & $5.23^{+0.45}_{-4.67}$ & $3.38^{+0.29}_{-3.02}$ & 0.84(35.2/42) \\
\hline
\end{tabular}}

\vspace{2mm}
\noindent\footnotesize
The X-ray spectral fitting results for X3, with parameters defined same as those in Table~\ref{tab:2}.$f_X$ is the observed full-band X-ray flux in units of $10^{-14}\,\mathrm{erg\,cm^{-2}\,s^{-1}}$,
and $L_X$ is the corresponding full-band X-ray luminosity in units of $10^{38}\,\mathrm{erg\,s^{-1}}$.
\end{table*}

\begin{table}[ht]
\centering
\caption{XMM-Newton Spectral Fitting Data of X4\label{tab:4}}
\small
\begin{tabular}{lcc}
\hline
Parameter & ObsID (110900201) & ObsID (0890710101) \\
\hline
 & MOS & MOS \\
$N_{\mathrm{H, Gal}}$ & (1.3) & (1.3) \\
$N_{\mathrm{H, 1}}$ & (3.2) & (3.2) \\
$N_{\mathrm{H, 2}}$ & $25.0^{+7.3}_{-15.9}$ & $11.2^{+9.8}_{-8.7}$ \\
$kT_{\mathrm{rs}}$ (keV) & (1.24) & (1.24) \\
$Z(Z_{\odot})$ & (1.0) & (1.0) \\
$K_{\mathrm{rs}}$ & $4.24^{+7.21}_{-4.24}\times 10^{-6}$ & $1.53^{+6.51}_{-1.53}\times 10^{-6}$ \\
$\Gamma$ & $1.75^{+0.22}_{-0.22}$ & $1.57^{+0.34}_{-0.41}$ \\
$N_{\mathrm{pl}}$ & $3.12^{+0.83}_{-0.71}\times 10^{-5}$ & $1.73^{+0.75}_{-0.68}\times 10^{-5}$ \\
$kT_{\mathrm{dbb}}$ (keV) & $0.19^{+0.03}_{-0.19}$ & (0.19) \\
$K_{\mathrm{dbb}}$ & (3.62) & (3.62) \\
\hline
$\chi^2/\mathrm{dof}$ & 1.04(56.2/54) & 1.01(59.3/59) \\
Flux & $2.16^{+0.16}_{-0.23}$ & $1.44^{+0.24}_{-0.24}$ \\
$L_{\mathrm{X}}$ & $1.40^{+0.10}_{-0.15}$ & $0.93^{+0.16}_{-0.16}$ \\
\hline
\end{tabular}

\vspace{2mm}
\noindent\footnotesize
The best-fit parameter values for the \textit{XMM-Newton} spectra of X4. 
Errors are 90\% confidence levels for one interesting parameter.
The spectral model contains an optically-thin plasma component (\texttt{raymond}), and the disk blackbody plus power law component (\texttt{powerlaw + diskbb}) component. 
$N_{\mathrm{H,Gal}}$ is the Galactic absorption column density, while $N_{\mathrm{H,1}}$ and $N_{\mathrm{H,2}}$ are the intrinsic absorption column densities associated with the \texttt{raymond} and the \texttt{powerlaw + diskbb} components, respectively, all in units of $10^{20}$ cm$^{-2}$.
$kT_{\mathrm{rs}}$ and $kT_{\mathrm{dbb}}$ are the temperatures of the \texttt{raymond} and \texttt{diskbb} components in keV, while $K_{\mathrm{rs}}$ and $K_{\mathrm{dbb}}$ are the normalizations of those two components, respectively. 
$\Gamma$ is the power-law photon index, while $N$ is the normalization of the power-law component. 
$Z(Z_{\odot})$ represents metallicity (fixed at 1.0).
$f_X$ is the observed full-band X-ray flux in units of $10^{-13}\,\mathrm{erg\,cm^{-2}\,s^{-1}}$,
while $L_X$ is the corresponding full-band X-ray luminosity in units of $10^{39}\,\mathrm{erg\,s^{-1}}$. 
\end{table}

\begin{table*}[ht]
\centering
\caption{Spectral fitting data of X5\label{tab:5}}
\resizebox{\textwidth}{!}{%
\begin{tabular}{ccccccccc}
\hline
Mission & obsID & $N_{\mathrm{H,Gal}}$ & $N_{\mathrm{H}}$ & $N_{\mathrm{pl}}$ & $\Gamma$ & Flux & $L_X$ & $\chi^2/\mathrm{dof}$ \\
\hline
Chandra  & 797   & 1.3 & $20.0^{+2.5}_{-2.4}$ & $0.90^{+0.08}_{-0.07}$ & $1.80^{+0.09}_{-0.08}$ & $4.18^{+0.16}_{-0.18}$ & $2.71^{+0.10}_{-0.12}$ & 0.95 (95.1/100) \\
Chandra  & 25220 & 1.3 & $59.3^{+71.5}_{-59.1}$ & $0.84^{+1.26}_{-0.48}$ & $2.40^{+0.75}_{-0.65}$ & $1.69^{+0.22}_{-0.98}$ & $1.09^{+0.14}_{-0.63}$ & 0.76 (15.1/20) \\
Chandra  & 25777 & 1.3 & $72.3^{+22.0}_{-19.6}$ & $3.54^{+1.15}_{-0.83}$ & $2.32^{+0.22}_{-0.21}$ & $7.18^{+0.41}_{-0.59}$ & $4.65^{+0.27}_{-0.38}$ & 1.26 (60.7/48) \\
Chandra  & 25778 & 1.3 & $105.5^{+76.9}_{-66.1}$ & $1.79^{+3.09}_{-1.07}$ & $2.89^{+0.84}_{-0.72}$ & $1.73^{+0.17}_{-0.91}$ & $1.12^{+0.11}_{-0.59}$ & 1.15 (27.6/24) \\
Chandra  & 25779 & 1.3 & $57.8^{+51.9}_{-45.4}$ & $0.94^{+0.87}_{-0.44}$ & $2.10^{+0.53}_{-0.48}$ & $2.58^{+0.29}_{-0.80}$ & $1.67^{+0.19}_{-0.52}$ & 0.62 (16.7/27) \\
Chandra  & 25781 & 1.3 & $29.5^{+72.5}_{-29.5}$ & $0.79^{+1.16}_{-0.36}$ & $2.22^{+0.72}_{-0.47}$ & $2.28^{+0.36}_{-1.47}$ & $1.48^{+0.23}_{-0.95}$ & 0.55 (9.3/17) \\
Chandra  & 25782 & 1.3 & $44.9^{+23.4}_{-21.0}$ & $1.96^{+0.66}_{-0.48}$ & $1.77^{+0.22}_{-0.21}$ & $9.54^{+0.67}_{-0.92}$ & $6.18^{+0.43}_{-0.60}$ & 1.10 (54.0/49) \\
Chandra  & 26485 & 1.3 & $13.6^{+62.2}_{-13.6}$ & $0.49^{+0.60}_{-0.17}$ & $1.73^{+0.64}_{-0.34}$ & $2.59^{+0.36}_{-1.43}$ & $1.68^{+0.23}_{-0.93}$ & 0.99 (14.8/15) \\
\hline
XMM(MOS) & 110900201 & 1.3 & $21.9^{+2.8}_{-2.6}$ & $1.28^{+0.11}_{-0.10}$ & $2.02^{+0.09}_{-0.08}$ & $6.04^{+0.23}_{-0.26}$ & $3.91^{+0.15}_{-0.17}$ & 0.67 (88.8/133) \\
XMM(PN)  & 110900201 & 1.3 & $25.7^{+2.2}_{-2.1}$ & $1.35^{+0.10}_{-0.10}$ & $2.22^{+0.08}_{-0.08}$ & $4.96^{+0.18}_{-0.18}$ & $3.21^{+0.12}_{-0.12}$ & 1.13 (100.5/89) \\
XMM(MOS) & 890710101 & 1.3 & $19.5^{+4.9}_{-4.4}$ & $0.96^{+0.15}_{-0.13}$ & $2.10^{+0.17}_{-0.15}$ & $4.31^{+0.30}_{-0.29}$ & $2.79^{+0.19}_{-0.19}$ & 0.72 (41.9/58) \\
XMM(PN)  & 890710101 & 1.3 & $22.1^{+3.8}_{-3.5}$ & $1.08^{+0.16}_{-0.13}$ & $2.28^{+0.18}_{-0.17}$ & $3.92^{+0.30}_{-0.27}$ & $2.54^{+0.19}_{-0.17}$ & 1.38 (84.3/61) \\
\hline
Swift & 82263001  & 1.3 & $2.12^{+30.4}_{-2.12}$ & $1.41^{+1.21}_{-0.38}$ & $1.50^{+0.57}_{-0.31}$ & $11.19^{+2.20}_{-4.66}$ & $7.24^{+1.42}_{-3.01}$ & 0.55 (6.1/11) \\
\hline
\end{tabular}}

\vspace{2mm}
\noindent\footnotesize
The X-ray spectral fitting results for X5. 
Errors are 90\% confidence levels for one interesting parameter.
$N_{\mathrm{H,Gal}}$ and $N_{\mathrm{H}}$ are the Galactic and intrinsic absorption column density in units of $10^{20}\,\mathrm{cm^{-2}}$, respectively. 
$\Gamma$ and $N_{\mathrm{pl}}$ are the photon index and the normalization of the \texttt{powerlaw} model, respectively. $N_{\mathrm{pl}}$ is given in units of $\ 10^{-4}\,\mathrm{photons\,keV^{-1}\,cm^{-2}\,s^{-1}}$ at 1 keV.
$f_X$ is the observed full-band X-ray flux in units of $10^{-13}\,\mathrm{erg\,cm^{-2}\,s^{-1}}$,
while $L_X$ is the corresponding full-band X-ray luminosity in units of $10^{39}\,\mathrm{erg\,s^{-1}}$. 
\end{table*}

\section{Results} \label{sec:4}
\subsection{Light Curves of X-ray Sources} \label{sec:4.1}



\begin{figure}[htbp]
\centering
\includegraphics[width=0.56\textwidth]{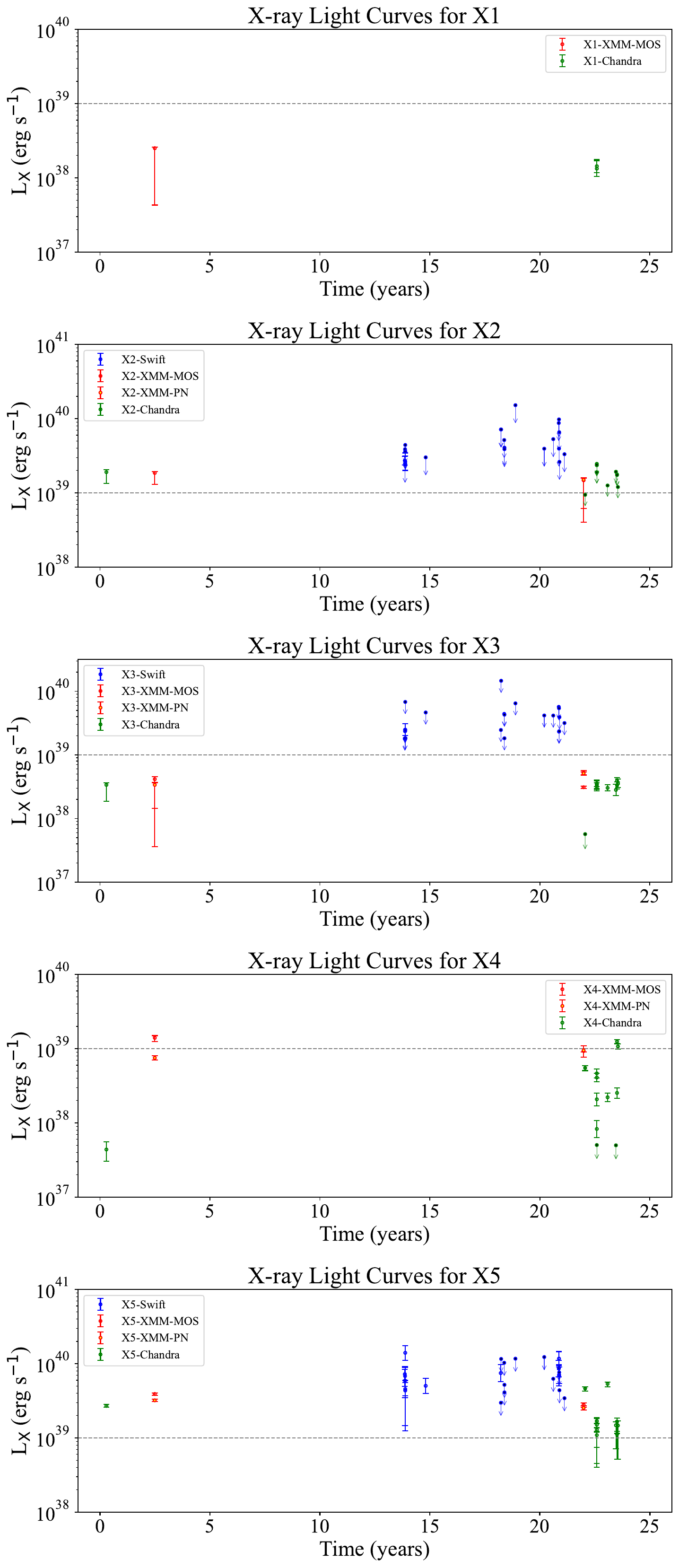}
\caption{X-ray light curves of X1--X5 constructed from \textit{Chandra}, \textit{XMM-Newton}, and \textit{Swift} observations. The vertical axis shows the full-band X-ray luminosity, while the horizontal axis indicates the time elapsed since the year 2000. Different colors denote the contributing missions (green: \textit{Chandra}; red: \textit{XMM-Newton}; blue: \textit{Swift}). Downward-pointing arrows indicate upper limits.}
\label{fig:lc1}
\end{figure}

The light curves for the five sources are constructed by combining the \textit{Chandra}, \textit{XMM-Newton}, and \textit{Swift} data, as shown in Figures~\ref{fig:lc1}.No \textit{Swift} flux measurements or upper limit values are reported for X1 or X4, because neither source is detected and no reliable constraints can be derived: X1 has a supersoft spectrum, while X4 lies too close to the bright neighboring source (X5) to allow robust upper limit estimates. When both the X-ray luminosity derived from spectral fitting and those calculated under the power-law assumption were available for the same observation, we adopt the spectral-fitting values preferentially. Upper limits are also included in the light curve, which are calculated assuming a Galactic-absorbed power law model, with photon index adopted as the average $\Gamma_{\rm eff}$ values from the observations  where the source is detected. 


The X-ray light curves covering 24 years  illustrate the X-ray variability behaviors of the five sources . While X3, X4, and X5 exhibit stronger variations by more than one order of magnitude. For X1, the X-ray luminosity values of the two \textit{Chandra} observations in 2022 , in which have this source detected are consistent with that in the 2002 \textit{XMM-Newton} observation during which X1 was in the high state \citep{soria2009different}. X1 had lower X-ray luminosity ($\approx10^{37}$~erg~s$^{-1}$) during its low state in the 2000 \textit{Chandra} observation, albeit with large uncertainties \citep{soria2009different}. X2 is a heavily absorbed ULX ($N_{\rm H} \sim 2 \times 10^{22}~{\rm cm^{-2}}$) with a “convex-spectrum” \citep{2007IAUS..238..209M}. Its X-ray spectra from multiple observations can be well fitted using the \texttt{diskbb} model, with a color temperature of  $kT_{\rm in} \sim 0.9$--$2.3~\mathrm{keV}$. Based on spectral fitting from \textit{Chandra} and \textit{XMM-Newton} observations, its absorbed luminosity remains around $L_{\rm X} \approx 2 \times 10^{39}~{\rm erg~s^{-1}}$in only a few observations. However,  strong absorption significantly affects the observed spectral shape in several \textit{Chandra} and \textit{Swift} observations , preventing a reliable spectral characterization and thereby inhibiting an accurate estimation of $\Gamma_{\rm eff}$ and the corresponding X-ray luminosity. In these cases, only upper limits are reported in the light curve. Source X3 reached the ULX luminosity threshold ($10^{39}~{\rm erg~s^{-1}}$) in a single \textit{Swift} observation (ObsID = 00082263005). Therefore, we also classify X3 as a ULX in NGC 4631.In this observation, the source underwent a brightening episode, with the full-band luminosity increasing to $\approx 2.5 \times 10^{39}\ \mathrm{erg\ s^{-1}}$, accompanied by a significant rise in the \texttt{diskbb} inner disk temperature to $kT_{\rm in} \sim 3\ \mathrm{keV}$. In the most recent \textit{Chandra} observations, its luminosity remained mostly stable at $L_{\rm X} \approx 4 \times 10^{38}~{\rm erg~s^{-1}}$, although one observation recorded a luminosity below $10^{38}~{\rm erg~s^{-1}}$. X4 exhibits large amplitude variability, with changes exceeding one order of magnitude within a few days. The X-ray spectra of X5 from \textit{Chandra}, \textit{XMM-Newton}, and \textit{Swift} observations can all be well described by a simple power-law model ($\Gamma \sim 2$). Its luminosity consistently remains high, never dropping below $10^{39}~{\rm erg~s^{-1}}$. In two \textit{Swift} observations it exceeded $10^{40}~{\rm erg~s^{-1}}$. 

\begin{figure}[htbp]
\centering
\includegraphics[width=0.65\textwidth]{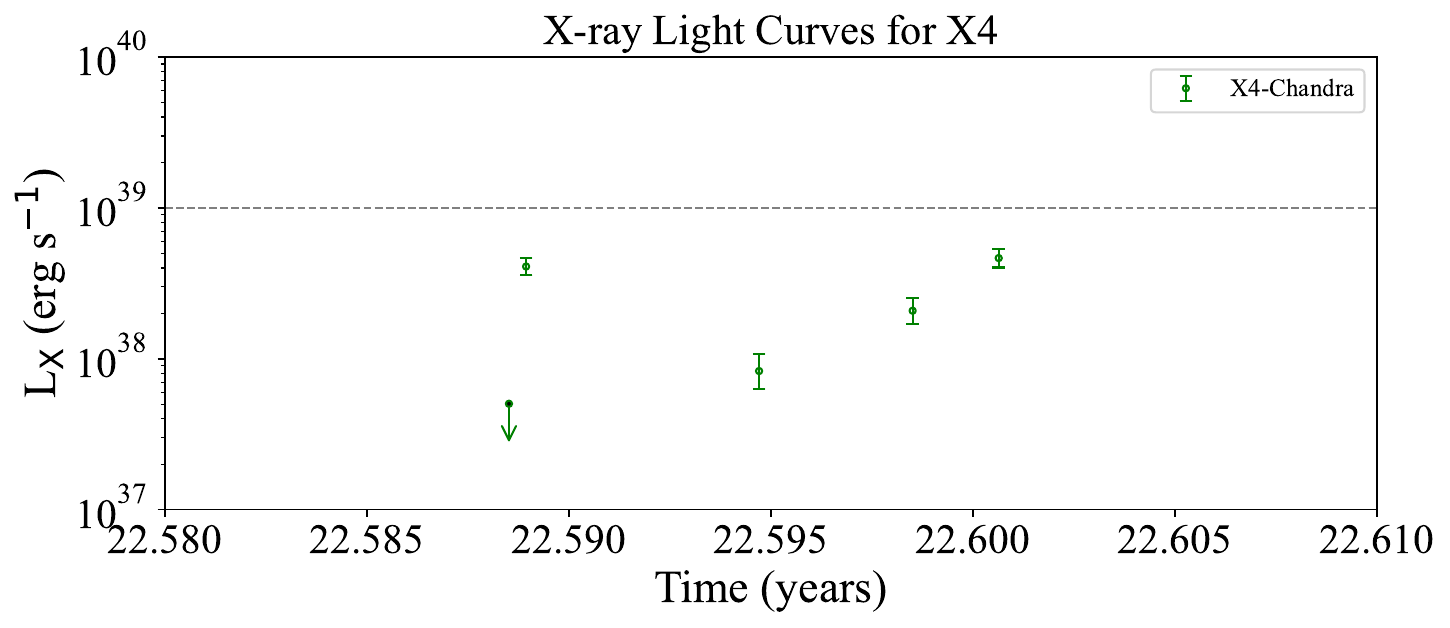}
\caption{The zoom-in X-ray light curves of X4. The figure follows the same labeling and annotation conventions as Figure~\ref{fig:lc1}.}
\label{fig:112}
\end{figure}
\begin{figure}[htbp]
\centering
\includegraphics[width=0.65\textwidth]{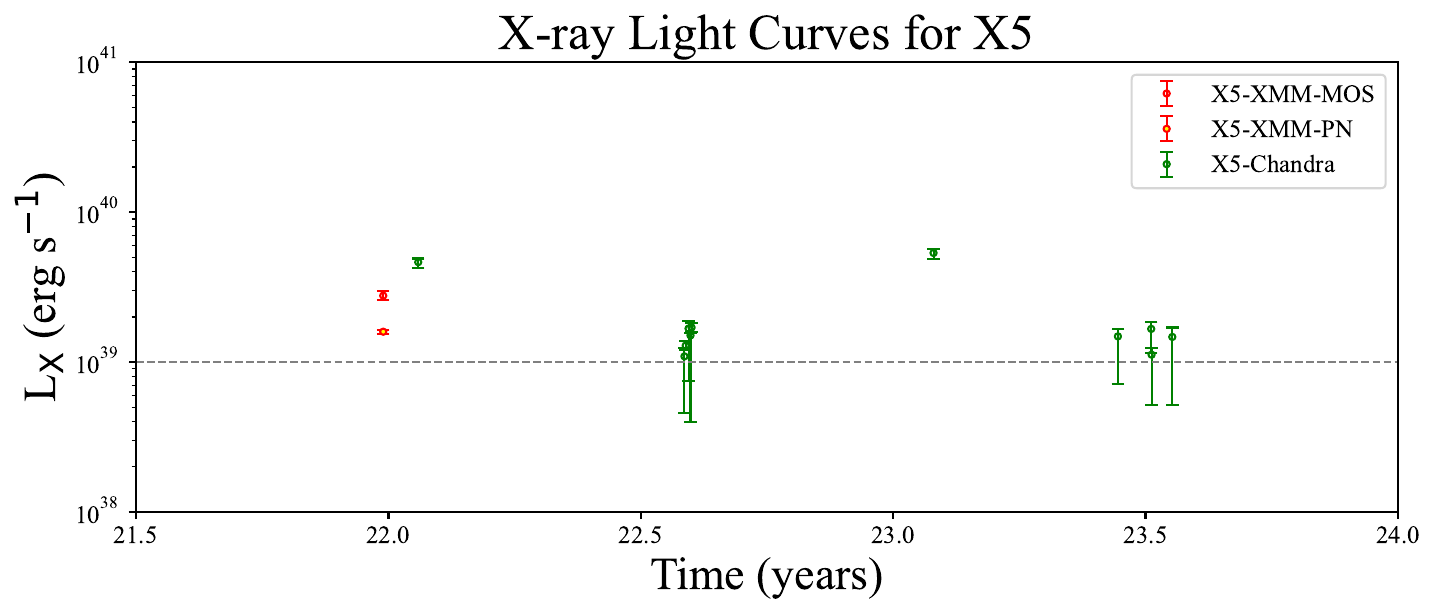}
\caption{The zoom-in X-ray light curves of X5 in the 21.5--24 year span, also following the same labeling and annotation conventions as Figure~\ref{fig:lc1}.}
\label{fig:111}
\end{figure}

\textit{Chandra} had relatively high-cadence observations on NGC 4631 during 2022 to 2023. We further present the zoom-in light curves during that period for X4 (Figure~\ref{fig:112}) and X5 (Figure~\ref{fig:111}) which have larger amplitude of variations. We find that the X-ray luminosity of X4 dropped by an order of magnitude within a single day and gradually returned to its previous level over the following three days. Using Lomb–Scargle periodogram analysis of the 11 \textit{Chandra} observations shown in Figure~\ref{fig:111}, the X-ray luminosity of X5 shows a possible $\sim311$-day modulation with flux variations of a factor of $\sim3$, although with low statistical significance due to the limited temporal coverage. However, the limited phase coverage of the current data prevents definitive conclusions. 

\subsection{Hardness Ratio of X-ray sources} \label{sec:4.2}

\begin{figure}[htbp]
\centering
\includegraphics[width=0.6\textwidth]{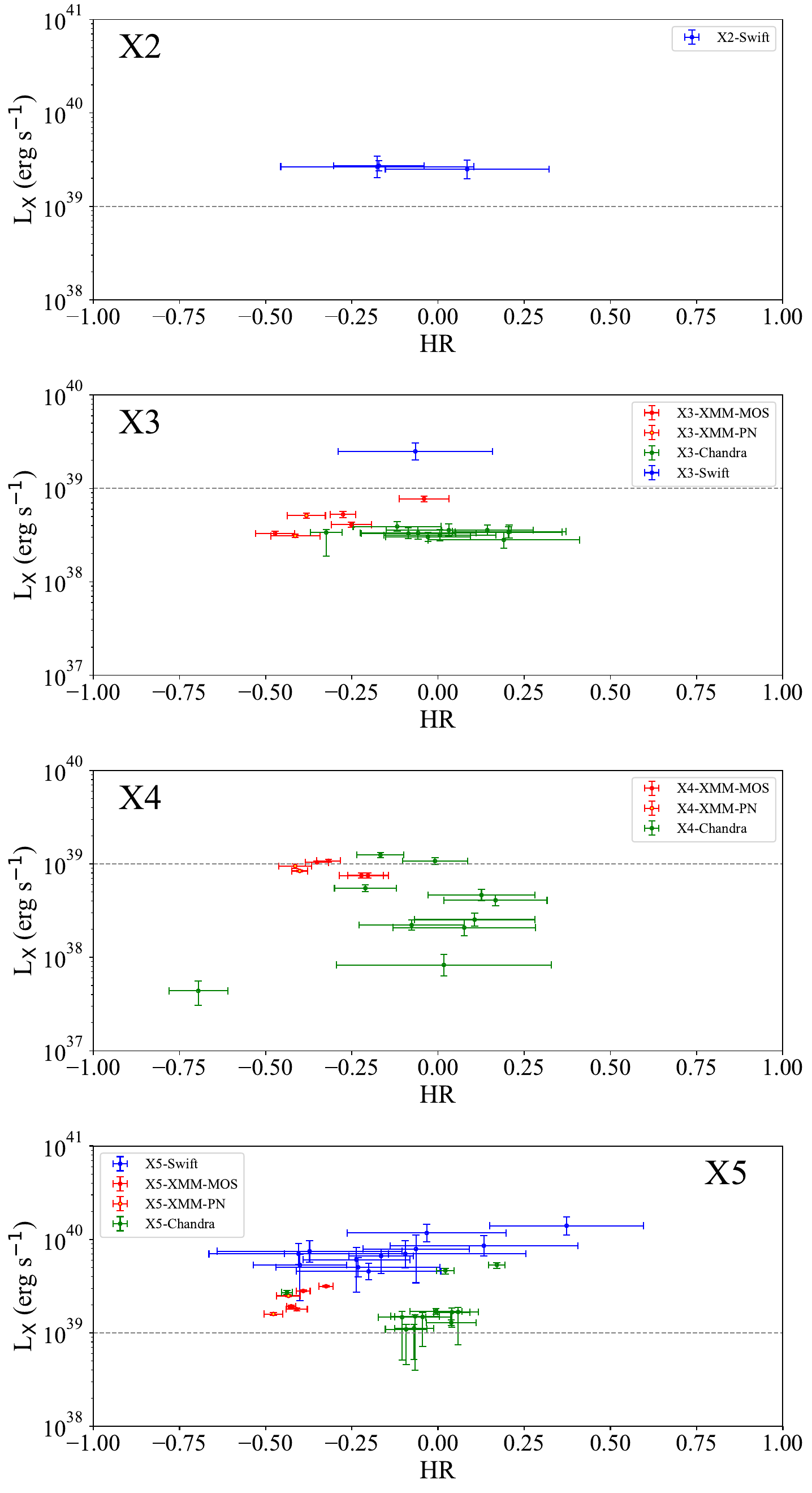}
\caption{The hardness ratio -- X-ray luminosity panels for X2--X5. The full-band X-ray luminosities $L_X$ are plotted versus hardness ratio $HR = (H-S)/(H+S)$, where $H$ and $S$ are the net counts in the hard and soft bands, respectively. Color and symbol annotations are the same as in Figure~\ref{fig:lc1}.}
\label{fig:lc5}
\end{figure}

We further investigate the relationship between X-ray luminosity and the hardness ratio ($HR$) which represent basic X-ray spectral properties. 
Based on the data in Tables~\ref{tab:flux2}–\ref{tab:flux5}, we construct luminosity versus $HR$ diagrams (Figure~\ref{fig:lc5}). This analysis helps determine whether the observed luminosity variations are primarily driven by enhancements in the soft or hard X-ray component, thereby providing clues to the underlying physical mechanisms. It should be noted that only $HR$ values from the same instrument can be compared. 

From the luminosity--hardness ratio diagrams, X5 exhibits a positive correlation in multiple observations with the same instrument, i.e., higher luminosity corresponds to a larger hardness ratio, indicating a harder spectrum.  To quantify this relationship, we computed the Pearson correlation coefficient $r$ and null-hypothesis probability $p$ for each instrument. For \textit{Swift} (epoch number $N$ = 12), we obtained $r = 0.77$ with $p = 0.003$, corresponding to a significance of $3.0\sigma$, indicating a clear positive correlation. For \textit{XMM} ($N$ = 6), $r = 0.83$ with $p = 0.041$ ($2.1\sigma$), also showing a positive trend, albeit marginal. However, the correlation analysis on \textit{Chandra} ($N$ = 12) data provides $r = 0.26$ with $p = 0.42$ ($0.8\sigma$), indicating no significant correlation.Three of \textit{Chandra} data points, which have substantially higher luminosity, appear to be disjoint from others, possibly indicating state transitions between the observations, although with very limited statistics. The outlying point with the lowest $HR$ value is from the observation of ID 797 (Cycle 1, ACIS-S), while all the remaining observations are obtained in Cycle 23 for which the ACIS-S detectors have had significant degradation in the quantum efficiency, especially in the soft energy band. Therefore, the pronounced difference in $HR$ value likely due to the instrumental response effects rather than intrinsic spectral softening. After excluding the Cycle 1 outlier, the Chandra sample (N = 11) yields a Pearson correlation coefficient of r = 0.70 with p = 0.017 ($2.4\sigma$) for X5, indicating improved positive correlation. In contrast, the other three sources X2, X3 and X4 show no obvious correlation between luminosity and hardness ratio, implying that their luminosity variations may not depend on spectral hardness, or their variability mechanisms are more complex. As X1 is a supersoft source, it was not detected in the hard X-ray band, and thus no hardness ratio diagram could be produced.



\subsection{Normalized Excess Variance} \label{sec:4.3}

To quantify the variability amplitude, we adopt the \textit{normalized excess variance} ($\sigma_{\rm rms}^2$) as a measure of the strength of luminosity variations, and further investigate its relationship with the X-ray luminosity. The normalized excess variance is a statistical quantity that characterizes the amplitude of intrinsic variability by considering the deviations of all observations relative to the mean luminosity, thereby reflecting the overall degree of variability. Following \citet{Lanzuisi_2014}, $\sigma_{\rm rms}^2$ and its associated uncertainty are defined as:
\begin{equation}
\sigma_{\rm rms}^2 = \frac{1}{(N_{\rm obs}-1)\bar{x}^2} \sum_{i=1}^{N_{\rm obs}} (x_i-\bar{x})^2 - \frac{1}{N_{\rm obs}\bar{x}^2} \sum_{i=1}^{N_{\rm obs}} \sigma_{{\rm err},i}^2
\label{eq:sigma_rms}
\end{equation}

\begin{equation}
{\rm err}(\sigma_{\rm rms}^{2}) = \sqrt{\left(\sqrt{\frac{2}{N_{\rm obs}}} \frac{\overline{\sigma_{\rm err}^{2}}}{\bar{x}^{2}}\right)^{2} + \left(\sqrt{\frac{\overline{\sigma_{\rm err}^{2}}}{N_{\rm obs}}} \frac{2 F_{\rm var}}{\bar{x}}\right)^{2}}
\label{eq:sigma_rms_err}
\end{equation}

where $N_{\rm obs}$ is the number of observations, $\bar{x}$ is the mean luminosity defined as
$\frac{1}{N_{\rm obs}} \sum_{i=1}^{N_{\rm obs}} x_i$,
 $x_i$ is the luminosity of the $i$-th observation, $\sigma_{{\rm err},i}$ is the measurement error of $x_i$, and $F_{\rm var} = \sqrt{\sigma_{\rm rms}^2}$.

\begin{figure}[htbp]
\centering
\includegraphics[width=0.5\textwidth]{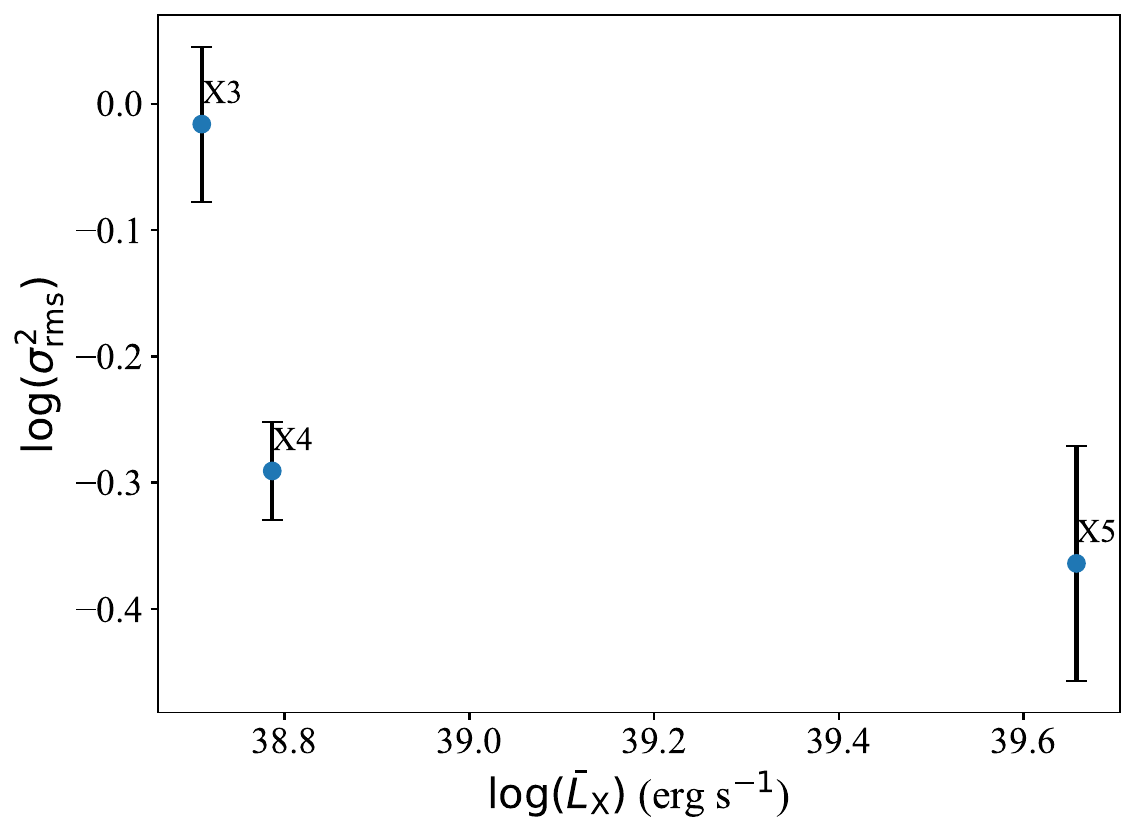}
\caption{The relation between mean X-ray luminosity in the full-band and the normalized excess variance ($\sigma_{\rm rms}^2$). The horizontal axis shows the mean full-band X-ray luminosity for each source.
}
\label{fig:lc7}
\end{figure}

The two terms on the right-hand side of Equation~\ref{eq:sigma_rms} represent: (1) the total variance of the observed data, accounting for the squared deviations of each observation from the mean, which includes both the intrinsic variability of the source and the observational uncertainties; (2) the contribution of measurement errors to the total variance.Both terms are normalized by $\bar{x}^2$, yielding a dimensionless measure of variability, without affecting their decomposition or physical interpretation. Subtracting the measurement error contribution yields a more accurate estimate of the intrinsic variability amplitude.
We calculated the $\sigma_{\rm rms}^2$ values for sources X3, X4, and X5 based on their observed luminosities, as shown in Figure~\ref{fig:lc7}. We do not compute the excess variance for X1 and X2 due to the limited number of reliable detections and the presence of a significant fraction of upper limits. Although the available data points are limited, it can be seen that $\sigma_{\rm rms}^2$ decreases with increasing luminosity, i.e., sources with higher  average luminosity in 0.5-8.0~keV exhibit smaller variability amplitudes on years timescales.. This behavior is consistent with the previously reported results in \citet{Gonz_lez_Mart_n_2011}), although they had been focused on much shorter timescales of kilo-seconds (i.e., within an X-ray observation) and on a different energy band 2.0-10~keV..


\subsection{Structure Function} \label{sec:sf}

To investigate the evolution of variability amplitude with timescale, we employ the \textit{structure function} (SF) to represent the characteristic timescale of X-ray variability for X3, X4, and X5. Because X1 (an ultra-soft source) and X2 (a heavily absorbed source) have poor spectral quality, which results in insufficient data points for a reliable SF analysis. The SF is a tool to describe how the variability amplitude changes with time interval and is particularly useful for analyzing the time dependence of variability. According to \citet{prokhorenko2024xrayvariabilitysdssquasars}, the SF is defined as:
\begin{equation}
\mathrm{SF}^2(\Delta t) = \left\langle \left[ \log \frac{F_{X, \rm true}(t_{\rm obs} + \Delta t_{\rm obs})}{F_{X, \rm true}(t_{\rm obs})} \right]^2 \right\rangle
\label{eq:sf_def}
\end{equation}
where $\langle \cdot \rangle$ denotes the ensemble average, $F_{X, \rm true}$ is the intrinsic flux at a given time in the absence of measurement noise, and $\Delta t$ represents the time interval between two observations. 

We first construct all possible data pairs $(i,j)$ from the observational time series $t_i$ and corresponding luminosities $L_i$, and compute the time intervals $\Delta t_{ij}$, keeping only pairs with $\Delta t_{ij}>0$. To study the dependence of SF on time interval, $\Delta t$ is binned in logarithmic space with an initial bin width of 0.1 dex. If a bin contains too few data points, adjacent bins are merged according to two criteria: (1) the merged bin contains at least 10 data points; (2) the merged bin width in logarithmic space is no less than 0.1 dex. This ensures that the SF in each bin has statistical significance and reduces biases from small sample sizes.

For each merged bin, we extract all $\Delta t_{ij}$ and $\mathrm{SF}^2_{ij}$ data points. To avoid asymmetries caused by extreme values, the SF is analyzed in logarithmic space: we calculate $\log(\mathrm{SF}^2)$, its mean $\langle \log(\mathrm{SF}^2) \rangle$, and the standard error of the mean (SEM = $\sigma/\sqrt{N}$), where $\sigma$ is the sample standard deviation of $\log(SF^2)$ within each bin, and $N$ is the number of points in the bin. The time interval for the bin is defined as the mean of $\log(\Delta t)$ converted back to linear space. This yields the binned statistics $(\Delta t_{\rm bin}, \mathrm{SF}^2_{\rm bin})$.

Assuming a power-law relation between SF and time interval  ,as commonly adopted \citet{prokhorenko2024xrayvariabilitysdssquasars}, we have:

\begin{equation}
\mathrm{SF}^2(\Delta t) = A^2 \cdot (\Delta t)^\gamma
\label{eq:sf_powerlaw}
\end{equation}
where $A$ is the normalization constant, representing the variability amplitude at a reference timescale, and $\gamma$ is the slope of the relation. Equation~\eqref{eq:sf_powerlaw} can be rewritten linearly as:
\begin{equation}
\log(\mathrm{SF}^2) = \log(A^2) + \gamma \cdot \log(\Delta t)
\label{eq:sf_linear}
\end{equation}

\begin{figure}[htbp]
\centering
\includegraphics[width=0.7\textwidth]{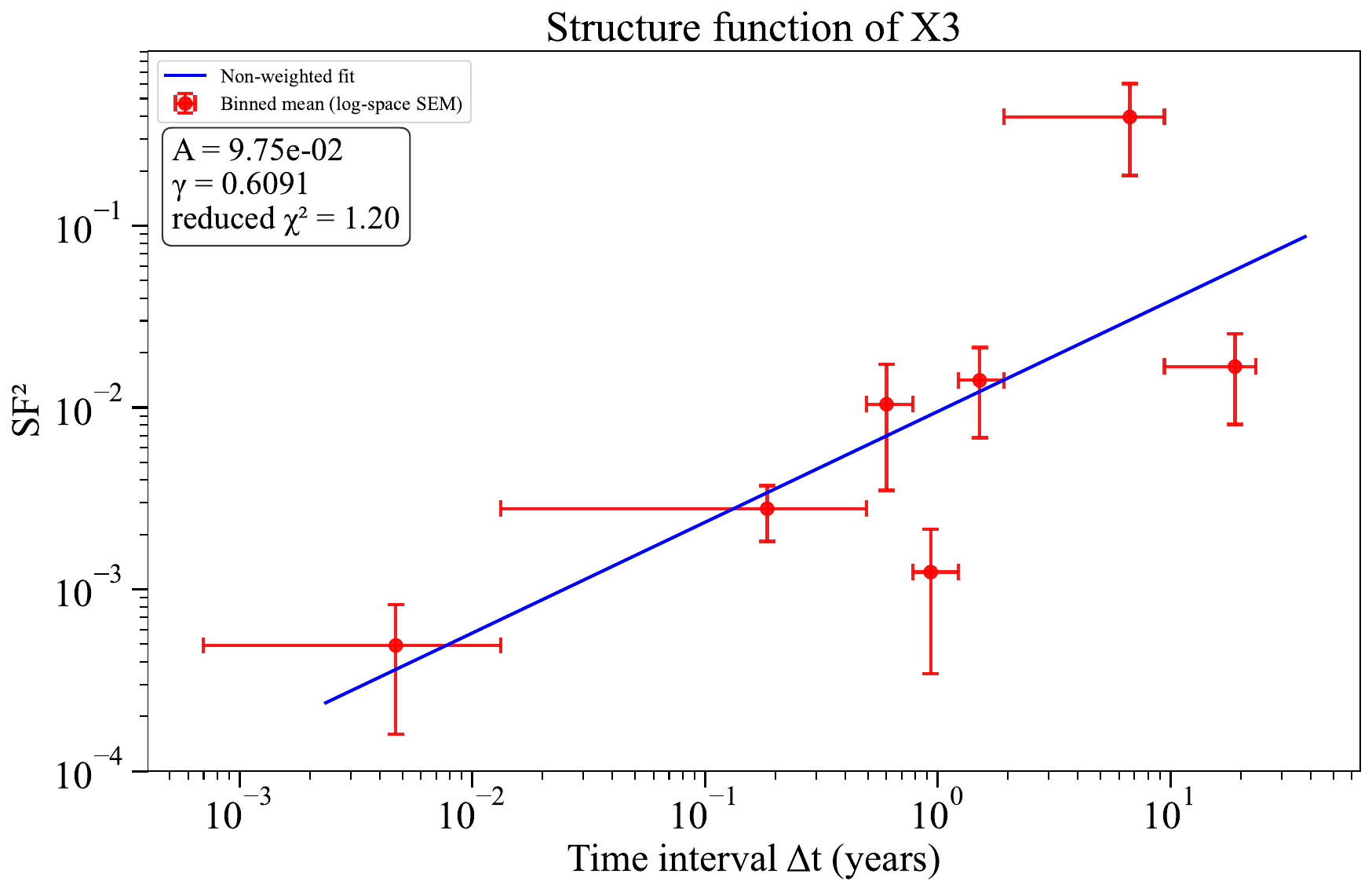}
\caption{The structure function ($\mathrm{SF}^{2}$) of X3. The plot uses logarithmic scales for both axes. The vertical axis shows the values of $\mathrm{SF}^{2}$, and the horizontal axis shows the time intervals between observations. The blue solid line shows the power-law fitting result for the two parameters, while the best-fit values are displayed in the upper-left corner of the figure.}
\label{fig:lc8}
\end{figure}

\begin{figure}[htbp]
\centering
\includegraphics[width=0.7\textwidth]{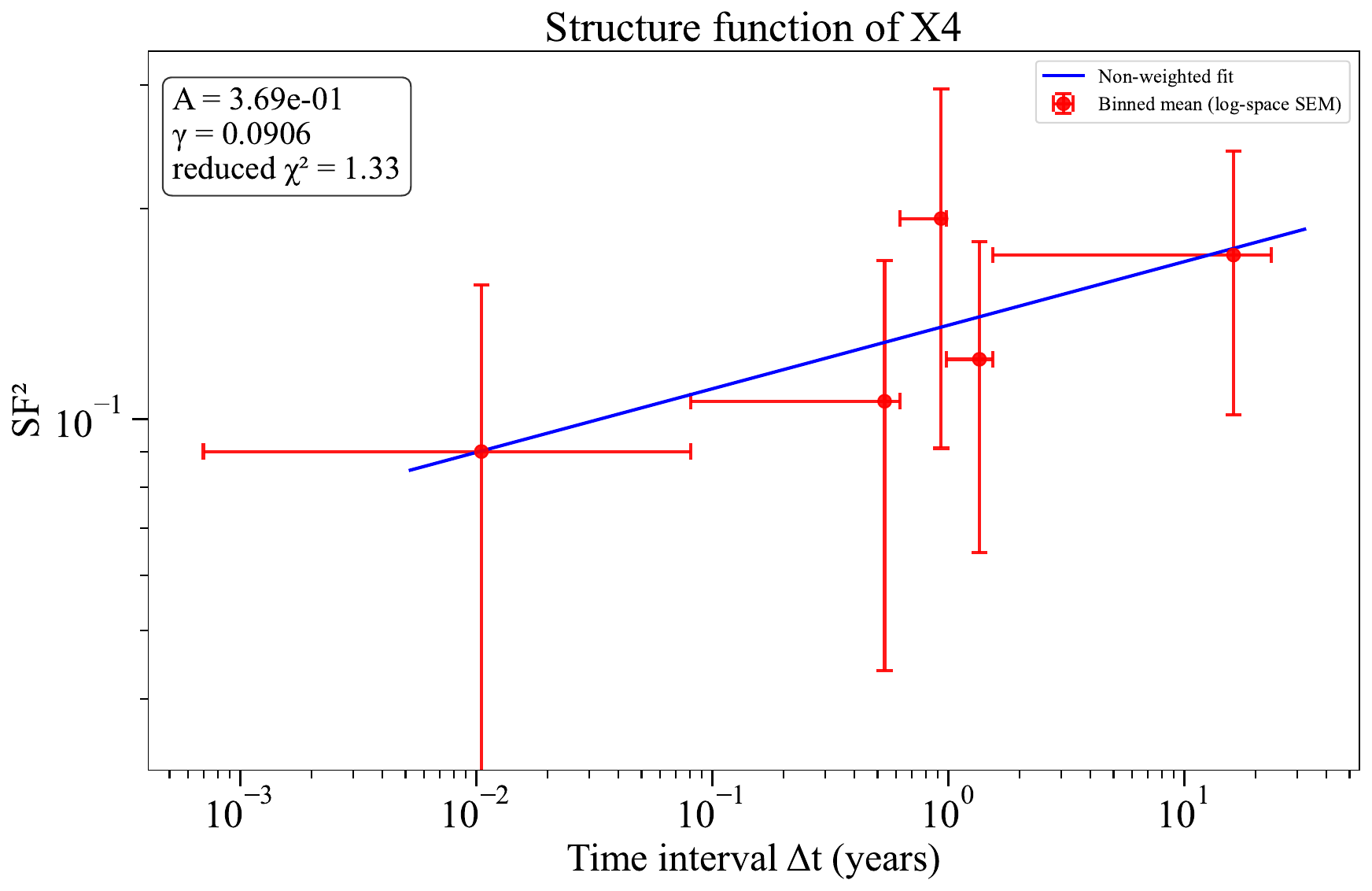}
\caption{The structure function ($\mathrm{SF}^{2}$) of X4. The figure caption is the same as that of Figure~\ref{fig:lc8}.}
\label{fig:lc9.}
\end{figure}

\begin{figure}[htbp]
\centering
\includegraphics[width=0.7\textwidth]{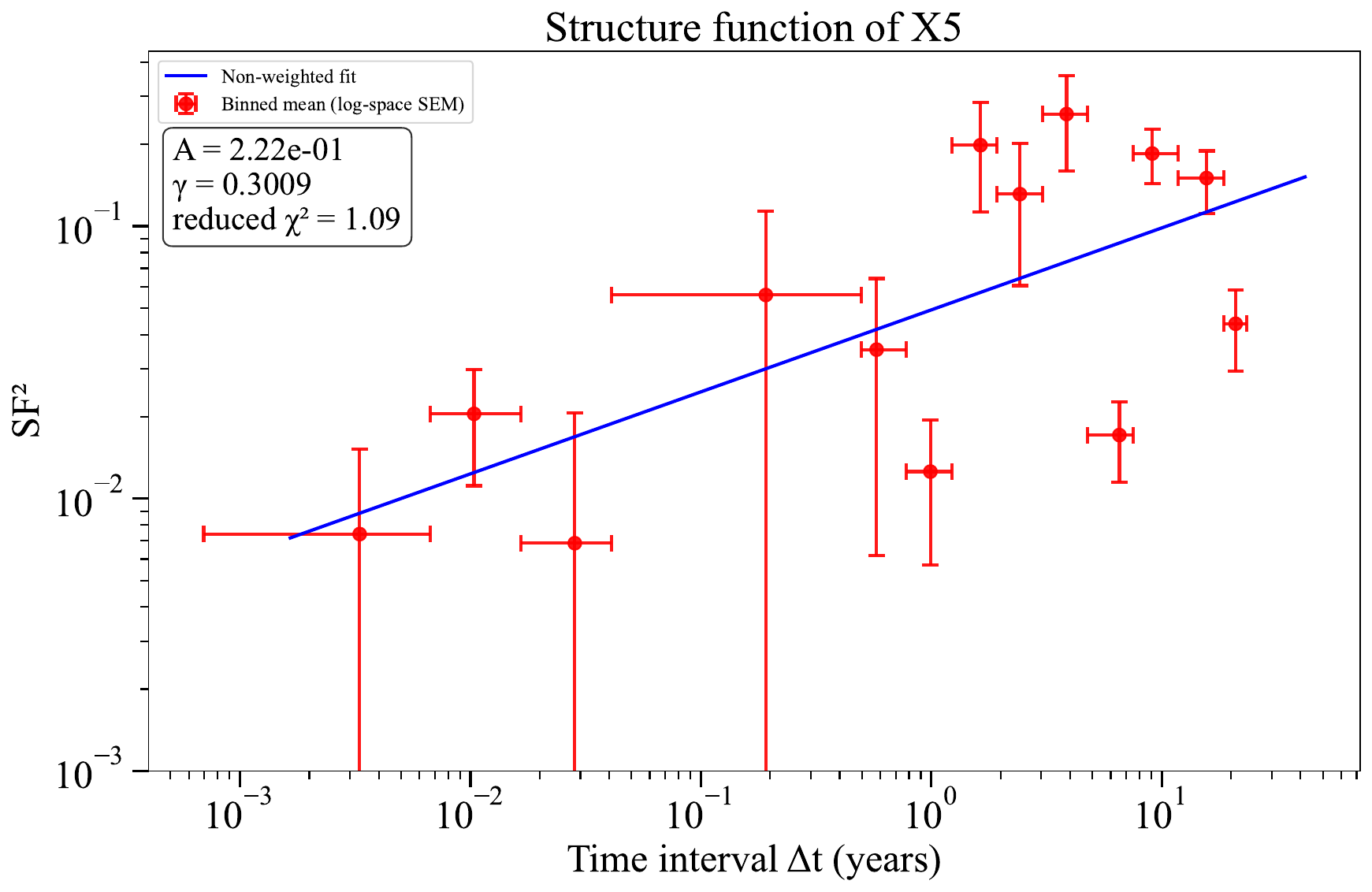}
\caption{The structure function ($\mathrm{SF}^{2}$) of X5. The figure caption is the same as that of Figure~\ref{fig:lc8}.}
\label{fig:lc10}
\end{figure}

We fit this linear form using an unweighted least-squares method to obtain the parameters $\log(A^2)$ and $\gamma$, along with their $1\sigma$ uncertainties. The goodness of fit is assessed by computing residuals and the reduced chi-square ($\chi^2_{\rm red}$). The $\mathrm{SF}^2$–$\Delta t$ relations for the three sources and the best-fit parameters for the power law are shown in Figure~\ref{fig:lc8}-\ref{fig:lc10}.

The structure function analysis shows that  $\mathrm{SF}^{2}$ generally increases with the time interval $\Delta t$ for X3 and X5, whereas X4 is consistent with a constant within the errors, indicating no significant variability trend. Among them, X3 exhibits a relatively small short-term (days to weeks) variability amplitude ($A \approx 9.75\times10^{-2}$) but the steepest power-law slope ($\gamma \approx 0.61 \pm 0.24$), implying that its variability increases more rapidly with increasing timescale, while remaining relatively weak on short timescales. In contrast, X4 shows a larger short-term variability amplitude ($A \approx 0.369$) with an almost flat slope ($\gamma \approx 0.09 \pm 0.05$), suggesting strong variability on short timescales but no significant increase in variability toward longer timescales. This behavior is consistent with the observed luminosity outbursts—rising by several factors within hours and then decaying within days (Figure~\ref{fig:112}). X5 has intermediate variability behavior between X3 and X4 from the perspective of structure function.

\section{Discussion} \label{subsec:discovery}

\subsection{Individual ULXs} \label{subsec:five_ulxs}

\subsubsection{X1} \label{subsubsec:x1}

Among the five brightest X-ray sources in NGC~4631, we performed a detailed analysis of source X1 and successfully reproduced the \textit{XMM-Newton} spectral fitting results obtained by \citet{soria2009different}.The 2000 Chandra observation (ObsID 797), corresponding to a low state of X1, was previously analyzed by \citet{soria2009different}, who fitted the spectrum with a disk-blackbody model under Cash statistics, obtaining an inner-disk temperature of $kT_{\rm in} \approx 0.2$ keV. We also re-analyzed this dataset using a disk-blackbody model with Cash statistics to account for the low photon statistics, finding results consistent with those reported in the literature.  During the subsequent $\sim$20~years of observations, X1 was detected only in two \textit{Chandra} exposures in 2022 (ObsIDs 25220 and 26484), suggesting that the source likely remained in a low state for most of this period. The 0.5–8~keV count rates in these two observations, both taken on the same day, were $(1.3 \pm 0.2)\times10^{-3}$ and $(1.1 \pm 0.2)\times10^{-3}$~counts~s$^{-1}$, respectively—about 3–4 times higher than that observed in 2000 (ObsID 797) when the source was in a low state—indicating that X1 underwent a brief rebrightening episode, possibly re-entering a high state. However, in the three following \textit{Chandra} observations over the next 3–6 days, X1 was not detected, with the count rate decreasing on average by about an order of magnitude, indicating strong variability over the timescale of days. 

To investigate the spectral properties of the two \textit{Chandra} detections (ObsIDs 25220 and 26484) in 2022, we fitted the  spectra  using an absorbed disk-blackbody model under Cash statistics. Although the limited photon statistics prevent detailed modeling,  both spectra are adequately described by this model, yielding inner-disk temperatures of $kT_{\mathrm{in}} = 0.15^{+0.05}_{-0.05}$~keV ($Cstat = 20.5/21$) and $kT_{\mathrm{in}} = 0.07^{+0.02}_{-0.02}$~keV ($Cstat = 8.6/12$), suggesting further spectral softening. Interestingly, despite the temperature decrease, the 0.5–8~keV count rate did not decline correspondingly, implying an increase in the total radiative power. This apparent ``cooling yet brightening'' behavior can be understood as a model-dependent effect: for the same observed count rate, a lower disk temperature shifts the spectral peak toward lower energies, thereby reducing the flux contribution within the instrumental bandpass. To reproduce the observed counts, the model requires a higher normalization, which, when extrapolated to the full energy range, yields a larger bolometric luminosity. Therefore, the 2022 spectral results support the interpretation that X1 was either entering or already in a high state at that time.

However, X1 was not detected in subsequent observations within the same year, implying that its high state was extremely short-lived. Such a ``rapid brightening–fading'' behavior is consistent with the ``fireball'' scenario proposed by \citet{soria2009different}, in which X1 is interpreted as a white dwarf undergoing a brief super-Eddington nuclear-burning episode. In this model, thermonuclear burning on the surface of the white dwarf causes the photosphere to expand dramatically to a radius of $\sim10^{9}$~cm, resulting in a drop in effective temperature but an increase in emitting area, maintaining a high luminosity of $L_{\mathrm{X}} \sim 10^{39}$~erg~s$^{-1}$. After the burning ceases, the photosphere contracts rapidly back to the stellar surface, the temperature rises again, and the X-ray flux declines sharply—manifesting as a transient high/soft state.

Alternatively, if X1 were interpreted as a super-Eddington accreting compact object, its extremely soft spectrum ($kT_{\mathrm{in}} < 0.15$~keV) and the very short duration of the high state would be inconsistent with the behavior of typical black hole or neutron star supercritical accretors. Combined with its high luminosity ($L \sim 10^{39}\,\mathrm{erg\,s^{-1}}$), such a low temperature 
($\sim0.1$~keV) implies, under the simple blackbody assumption following the temperature–luminosity relation $L = 4\pi R^2 \sigma T^4$, an emitting radius of approximately $R_{\rm BB} \approx 10^9$~cm. This value is far larger than the characteristic inner-disk scales around a black hole, but comparable to the photospheric radius of a white dwarf. Moreover, the high state persisted for only a few days—consistent with the timescale of a nova-like thermonuclear outburst on a white dwarf surface, but much shorter than that of a super-Eddington phase in X-ray binaries. Its peak luminosity ($\sim10^{39}$~erg~s$^{-1}$), though slightly exceeding the Eddington limit of a white dwarf, remains physically plausible when accounting for anisotropic emission or transient super-Eddington expansion effects.

\subsubsection{X2} \label{subsubsec:x2}

X2 consistently exhibits strong and persistent absorption, with $N_{\mathrm{H}} \approx 2\times10^{22}~\mathrm{cm^{-2}}$. Such long-term high absorption is likely associated with the dense environment of a young stellar cluster or molecular cloud \citep{soria2009different,Guo__2023}. The stable, elevated $N_{\mathrm{H}}$ introduces uncertainties in the derived blackbody temperature and bolometric luminosity, emphasizing the need for further constraints on the absorber’s spatial extent, ionization state, and kinematic properties to better understand the physical nature of X2.

In \textit{Swift} observations the source shows comparable numbers of soft and hard-band counts, resulting in a different count distribution compared to the \textit{XMM-Newton} and \textit{Chandra} data. This discrepancy is likely due to a combination of limited photon statistics and the lower sensitivity of \textit{Swift}, which can bias the hardness ratio in low-count regimes. In contrast, the higher-quality \textit{XMM-Newton} and \textit{Chandra} observations, with significantly better photon statistics, provide a more reliable characterization of the source emission and consistently indicate hard spectral behaviors. On the other hand, a possible contribution from variable absorption associated with ULX-driven winds cannot be completely excluded, as changes in the geometry or in the fraction of the source covered by the outflowing material along the line of sight may enhance the apparent soft component and modify the observed spectral shape.

\subsubsection{X3} \label{subsubsec:x3}

The X-ray spectrum of X3 can be well fitted with a multicolor disk model (\texttt{diskbb}) in multiple {\it Chandra} and {\it XMM-Newton} observations, with a typical inner disk temperature of $\approx$1.1--1.4 keV and a full-band X-ray luminosity of approximately $\approx$(3--5)$\times 10^{38}$ erg s$^{-1}$. However, in one {\it Swift} observation in 2018, the source underwent a brightening phase, with the full-band luminosity increasing to $\approx2.5\times 10^{39}$ erg s$^{-1}$, accompanied by a significant rise in the fitted inner disk temperature to $\sim3$~keV. Due to the limited photon statistics, the uncertainties on the fitted parameters are relatively large, but the trend of temperature increase remains evident, suggesting that this high-luminosity event may be associated with a substantial change in the structure of the accretion flow.

Assuming that this event still follows the standard thin-disk model and adopting a typical color correction factor of 1.7--2.0 \citep{Gierlinski_2004,Shafee_2006}, the inferred black hole mass would be significantly lower than the 5--7 $M_\odot$ estimated by \citet{soria2009different}. This indicates that during the high-luminosity episode, the standard thin-disk assumption may no longer hold, and the temperature in the \texttt{diskbb} model may no longer directly represent the effective temperature of the inner disk. Possible physical mechanisms leading to an apparently elevated temperature include, but are not limited to, strong Comptonization or the presence of a scattering layer, and supercritical accretion producing a thick disk with winds or funnel-like beaming. Notably, in a low-luminosity {\it Chandra} observation, the source luminosity dropped by approximately an order of magnitude, but the spectral hardness remained nearly constant. This suggests that the luminosity variation may not be caused by a change in the accretion state, but rather by geometric effects or the variations in non-disk components such as the power-law corona emission.

On the other hand, a neutron star accretor cannot be excluded for X3, since super-Eddington neutron star ULXs are known to reach similar luminosities and can exhibit disk-like or Comptonized spectra with comparable temperatures. However, in the absence of pulsation detections or other neutron-star-specific signatures, the current data do not allow us to distinguish between a neutron star and a black hole accretor.

\subsubsection{X4} \label{subsubsec:x4}

X4 is a transient ULX exhibiting state transitions which can be interpreted within the framework of the ``outer thermal disk + inner Comptonized/scattering region'' model proposed by \citet{soria2009different}. In this scenario, the outer standard thin disk dominates the low-energy thermal emission, while the innermost region near the black hole consists of a hot, optically thin corona that produces a Comptonized, power-law–like high-energy component. Recent \textit{Chandra} observations show that the luminosity of X4 increased rapidly from $\sim5\times10^{37}~\mathrm{erg~s^{-1}}$ to $\sim5\times10^{38}~\mathrm{erg~s^{-1}}$ within a few hours, and subsequently went through a declining and recovering process over several days (see Figure~\ref{fig:112} and Section~\ref{sec:sf}). Such variability likely originates from rapid adjustments in the Comptonizing region rather than large-scale structural changes in the accretion disk.

X4 may also be powered by a neutron star accretor, since super-Eddington neutron star ULXs can produce similar spectral shapes and exhibit rapid luminosity variability driven by changes in accretion geometry and beaming effects. In particular, variability with the timescale of hours are observed in confirmed pulsating ULXs (e.g., \citealt{Robba_2021}), which can be naturally explained by instabilities in the accretion column and variations in the beaming funnel. However, in the absence of coherent pulsations or other neutron-star-specific signatures, the nature of the compact object cannot be uniquely determined. 

\subsubsection{X5} \label{subsubsec:x5}

X5 has remained persistently bright throughout all observations, with X-ray luminosities consistently exceeding $10^{39}~\mathrm{erg~s^{-1}}$. Its spectrum is well described by a simple power-law model with a photon index of $\Gamma \approx 2$, without detectable soft thermal components or high-energy cutoffs. Our analysis reveals a positive correlation between the X-ray luminosity and spectral hardness (HR), i.e., the spectrum becomes slightly harder as the luminosity increases.A similar luminosity–hardness correlation has been reported in several ULXs (e.g., \citealt{luangtip2016x}), suggesting that this behavior is not uncommon among super-Eddington accretors. These sources typically exhibit multi-component spectra, consisting of a soft thermal component and a harder Comptonized component (e.g., \citealt{Gladstone_2009}). Within the framework of super-Eddington accretion, ULX spectral evolution can be interpreted in terms of a funnel-like accretion geometry formed by radiatively driven winds, where the observer views the system along the funnel axis. As the accretion rate increases, the wind photosphere expands and the funnel opening angle becomes narrower, leading to stronger geometric beaming and the increasing dominance of the hard emission component.

In contrast, X5 exhibits a power-law spectrum with no evidence for a significant soft thermal component. This luminosity–hardness relation likely reflects changes in the mass accretion rate and associated physical processes: as the accretion rate increases, the surrounding Comptonizing electron cloud becomes more active, scattering a larger fraction of low-energy photons to higher energies, thereby enhancing the hard X-ray contribution and causing mild spectral hardening\citep{kubota2002another}. The luminosity variations of X5 are therefore more likely driven by accretion rate fluctuations rather than by transitions between spectral states.
Its stable, power-law–dominated spectrum and mild luminosity–hardness correlation are consistent with the behavior of canonical ``power-law" type ULXs(e.g., the two ULXs in IC~342; \citealt{kubota2001observational}).

\subsection{Comparison between ULXs and AGNs} \label{subsec:agn_ulx}
By comparing the variability properties of ULXs and AGNs, we investigate whether similar ''variability–luminosity'' trends may be present in both classes. Previous studies have established that in AGNs, the normalized excess variance ($\sigma_{\mathrm{rms}}^{2}$) shows an inverse correlation with X-ray luminosity\citep{nandra1997dependence,turner1999x}. Applying the same analysis to ULX samples, \citet{Gonz_lez_Mart_n_2011} found that in the 2--10~keV energy band, the variability amplitude of ULXs also decreases with increasing luminosity, revealing a ``variability--luminosity'' anti-correlation analogous to that seen in AGNs.In our sample of three ULXs, we find a similar qualitative tendency, although the limited sample size precludes a statistically robust assessment of the relationship.

Moreover, structure function studies on AGNs have shown that the variability amplitude tends to increase with the length of the timescale, and is inversely correlated with luminosity, black hole mass, and accretion rate (e.g., \citealt{prokhorenko2024xrayvariabilitysdssquasars}). Our long-term monitoring of ULXs demonstrates a qualitatively similar trend  in two sources: the power-law slopes between structure function $\mathrm{SF}^2$ and timescale $\Delta t$ show that overall variability amplitude increases with longer temporal baselines. If confirmed for a larger sample of ULXs, this consistency may suggests that the physical processes governing X-ray variability in ULXs and AGNs may share a common underlying mechanism. If  a fraction of the ULX population are indeed scaled-down analogs of AGNs, then their observed variability properties should follow comparable scaling relations. 
The present results already exhibit signs of such a ``scaled extension'' of AGN variability behavior into the ULX regime, providing a potential avenue to estimate black hole masses and accretion rates in ULXs using  structure function $\mathrm{SF}^2$ as a diagnostic. Future observations with larger samples and longer temporal coverage will be crucial to confirm whether the relations between the variability amplitude and luminosity or timescale hold universally. 

\section{Summary} \label{sec:summary}

In this work, we present a systematic study on the long-term X-ray variabilities of the five ULXs (X1--X5) in the NGC 4631 galaxy which is covered by 37 epochs of X-ray observations spanning over two decades. X-ray light curves, spectral modeling, and statistical variability analyses are carried out to characterize the physical nature and accretion state of each source. The five ULXs reveal strikingly distinct physical characteristics. X1 is a supersoft source whose spectral properties are consistent with brief super-Eddington nuclear-burning episodes on the suface of a white dwarf, while X2 suffers persistent strong absorption, likely embedded within a dense stellar or molecular environment. Both of these two sources are detected in very limited number of epochs, precluding meaningful variability statistics. X3 is a canonical multicolor-blackbody ULX that undergoes occasional high-luminosity outbursts, during which it reaches the ULX luminosity threshold. X4 is a transient power-law ULX exhibiting rapid short-term flux variability, which can be attributed to energy redistribution in the inner Comptonizing region. X5 stands out as persistently luminous with a simple power-law spectrum and a marginal positive luminosity--hardness correlation ($\gtrsim2\sigma$), together with a possible annual variability cycle.

Statistical analyses on ULX long-term X-ray variabilities are conducted using X3, X4, X5 as a sample, with the following results:
\begin{enumerate}
    \item The normalized excess variance ($\sigma^2_{\rm rms}$) is inversely correlated with mean X-ray luminosity across these three sources, confirming that higher-luminosity ULXs exhibit smaller amplitude of flux fluctuations. 
    \item  The structure function (SF$^2$) reveals that X3 and X5 generally have more pronounced variability on the long timescale than on the short term, while X4 shows the opposite behavior.
\end{enumerate}

Strikingly, these behaviors closely parallel what have been observed in AGNs, where $\sigma^2_{\rm rms}$ anti-correlates with luminosity and variability amplitude grows with timescale. This suggests that at least a fraction of the ULX population shares the same underlying accretion variability physics as AGN. Future studies should expand the ULX sample and extend the temporal baseline with high-sensitivity X-ray monitoring to test the universality of the $\sigma^2_{\rm rms}$--$L_X$ anti-correlation and the $\mathrm{SF}^2$ time accumulation effect. Multiwavelength observations can further constrain the accretion environment and potential wind/jet activity of ULXs. Comparative studies of ULX variability with AGNs across different mass scales may offer quantitative tools for estimating black hole masses, inferring accretion rates, and understanding the universality of X-ray variability mechanisms, advancing our understanding of the ULX--AGN analogy.

\begin{acknowledgements}
This work is supported by the National Key R\&D Program of China under grants 2023YFA1607904, and the National Natural Science Foundation of China under grants 12273029 and 12221003.
\end{acknowledgements}

\bibliographystyle{raa}
\bibliography{references}

\end{document}